\documentclass[a4paper,11pt]{article}
\pdfoutput=1 
\usepackage{jheppub}	
\usepackage[utf8]{inputenc}
\usepackage[english]{babel}
\usepackage{makeidx}
\usepackage{amsfonts}
\usepackage{enumerate}
\usepackage{mathrsfs}
\usepackage{tensor}
\usepackage[autostyle]{csquotes}
\usepackage{subfig}
\newcommand{\ft}[2]{{\textstyle\frac{#1}{#2}}}

\newdimen\tableauside\tableauside=1.0ex
\newdimen\tableaurule\tableaurule=0.4pt
\newdimen\tableaustep
\def\phantomhrule#1{\hbox{\vbox to0pt{\hrule height\tableaurule
width#1\vss}}}
\def\phantomvrule#1{\vbox{\hbox to0pt{\vrule width\tableaurule
height#1\hss}}}
\def\sqr{\vbox{%
  \phantomhrule\tableaustep
\hbox{\phantomvrule\tableaustep\kern\tableaustep\phantomvrule\tableaustep}%
  \hbox{\vbox{\phantomhrule\tableauside}\kern-\tableaurule}}}
\def\squares#1{\hbox{\count0=#1\noindent\loop\sqr
  \advance\count0 by-1 \ifnum\count0>0\repeat}}
\def\tableau#1{\vcenter{\offinterlineskip
  \tableaustep=\tableauside\advance\tableaustep by-\tableaurule
  \kern\normallineskip\hbox
    {\kern\normallineskip\vbox
      {\gettableau#1 0 }%
     \kern\normallineskip\kern\tableaurule}%
  \kern\normallineskip\kern\tableaurule}}
\def\gettableau#1 {\ifnum#1=0\let\next=\null\else
  \squares{#1}\let\next=\gettableau\fi\next}
  
\def\apjl{\ref@jnl{ApJ}} 

\def\be{\begin{equation}}
\def\ee{\end{equation}}
\def\bea{\begin{eqnarray}}
\def\eea{\end{eqnarray}}
\newcommand{\nn}{\nonumber}

\title{ Light rings of five-dimensional geometries }

\author[a,c]{M. Bianchi,}
\author[b]{D. Consoli,}
\author[a,c]{A. Grillo,}
\author[c]{J.F. Morales}

\affiliation[a]{Dipartimento  di  Fisica,  Universit\`a  di  Roma  ``Tor  Vergata'' }
\affiliation[b]{Department of physics,  University of Wien}
\affiliation[c]{I.N.F.N.  Sezione  di  Roma  ``Tor  Vergata'', Via della Ricerca Scientifica, 00133 Roma, Italy}
\emailAdd{bianchi@roma2.infn.it}
\emailAdd{dario.consoli@univie.ac.at}
\emailAdd{morales@roma2.infn.it}
\emailAdd{alfredo.grillo@roma2.infn.it}

\abstract{We study massless geodesics near the photon-spheres of a large family of solutions of Einstein-Maxwell theory in five dimensions, including BHs, naked singularities and smooth horizon-less JMaRT geometries obtained as six-dimensional uplifts of  the five-dimensional solution.  
We find that a light ring of unstable photon orbits surrounding the mass center is always present, independently of the existence of a horizon or singularity.  We compute the Lyapunov exponent, characterizing the chaotic behaviour of geodesics near the `photon-sphere' and the time decay of ring-down modes dominating the response of the geometry to perturbations at late times. We show that,  for geometries free of naked singularities, the Lyapunov exponent is always bounded by its value for a Schwarzschild BH of the same mass.}

\preprint{PREPRINT}
 \clearpage

\begin{document}
\maketitle


\section{Introduction}
Starting from the seminal work by Strominger and Vafa \cite{Strominger:1996sh} on the microscopic origin of the BH entropy in five dimensions, 5-d BH's have become a very interesting play-ground where to explore ideas and develop tools finally aimed at the more realistic but by far more involved 4-d case. In String Theory, `large' (BPS) BH's with a finite area of the horizon require bound-states of strings and branes with at least three charges \cite{Cvetic:1995uj}. 
The three abelian charges couple to as many (gravi)photons, that in turn are dual to anti-symmetric tensors in five-dimensions, and to scalar fields parametrizing the shape and sizes of the internal manifold. One can `identify' the three charges so as to reduce the number of `active' (gravi)photons to one. In this setting scalar fields become constant and the relevant degrees of freedom are those of five-dimensional (super-)gravity coupled to a single gauge field {\it i.e.} Einstein-Maxwell theory. Thanks to the reduced complexity of the problem one can address the extension to the rotating case. The most general solution of this theory with three Killing symmetries was obtained by Chung, Cvetic, Lu and Pope (CCLP) in \cite{Chong_2005}. It depends on four parameters: mass ${\cal M}$, charge ${\cal Q}$ and two angular momenta $\ell_1$, $\ell_2$.
   
  The CCLP family includes BH solutions where the singularity is hidden behind an event horizon, over-rotating geometries and naked singularities where the singularity is unveiled. Naked singularities are ruled out by the cosmic censorship conjecture that excludes the existence of an accessible curvature singularity \cite{PhysRevLett.14.57,Penrose:1969pc}. Still, in a gravity theory with matter, an apparent singularity can be smoothed out by the blow-ups of gauge or scalar fields at the singularity signaling a break-down of the effective low-dimensional description and the emerging of extra-dimensions. An explicit realization of this mechanism is achieved by the so called JMaRT (after Jejjala, Madden, Ross and Titchener)family of solutions \cite{JMaRT}, that provide smooth horizon-less six-dimensional lifts of some over-rotating geometries. 
   
 It is natural to ask whether a more fundamental and unifying characterization of the solutions exists, where the details of the five-dimensional geometry and the existence or absence of horizons play a secondary role.  In this paper we perform this task and consider black holes (BH's), singular spaces and horizon-less geometries as a the target of gravitational scattering of massless neutral scalar particles in the vicinity of the photon-spheres \cite{Virbhadra:1999nm,Claudel:2000yi,Virbhadra:2002ju}. The photon-sphere of a BH or a very massive object is defined as the surface delimiting the scattering and absorption regions. It is visible as a light-ring of photons orbiting  around the mass distribution along `circular' orbits.  For a distant observer it appears as a shadow, whose rim is the projection on the celestial squashed sphere of the critical impact parameters $b^c_\theta$, $b^c_\phi$ (and in five-dimensions $b^c_\psi$).

The shape and size of the light ring provide a detailed imprint of the gravity solution that can eventually discriminate solutions with identical asymptotics (charges). The light ring determines also  the frequencies $\omega_{\rm QNM}$ of the so called quasi normal modes (QNM's) dominating at late times the gravitational wave response of the geometry to small perturbations  \cite{Vishveshwara:1970cc,Press:1971wr, 1972ApJ...172L..95G, PhysRevD.31.290, Ferrari:1984zz,Bombelli:1991eg,Cardoso:2008bp}.  
Our focus here is on the the Lyapunov exponent $\lambda$, governing the exponential deviation of nearby geodesics and setting the instability time scale for the decay of the basic QNM tone.  Indeed in the semi-classical approximation the complex frequencies of the quasi-normal modes are given by $\omega_{\rm QNM}= \omega_R(\ell) - i (2n+1) \lambda$ \cite{Berti:2009kk, Kokkotas:1999bd}, where the real part $\omega_R(\ell)$ is related to the frequency of the orbital motion with angular momentum $\ell$ and $n$ is known as the over-tone number. 

In \cite{Bianchi:2020des}, we found a bound $\lambda \le \lambda_{\rm Schw} =\sqrt{d-3}/(2b_{\rm Schw}) $ for the Lyapunov exponent in terms of the critical impact parameter $b_{\rm Schw}$ (the typical target size) of a Schwarzschild BH of the same mass\footnote{We also argued that the bound $\lambda \le \lambda_T= 2\pi T_{BH}$ found in \cite{Maldacena:2015waa} can be violated near extremality for charged and/or rotating BH's.}. We also provided some evidence that $\lambda$ can be used to discriminate a BH from its smooth horizonless micro-states.  Aim of this paper is to extend the analysis to the general CCLP solution of Einstein-Maxwell theory  in five dimensions. We will compute the Lyapunov exponent for the general four-parameter family and provide strong evidence that the bound is respected for BHs, regular and over rotating geometries where the singularity is hidden behind a horizon. Interestingly, the bound is violated at the internal photon-sphere of  spaces exposing a naked singularity, whose existence is ruled out by the cosmic censorship hypothesis \cite{Penrose:1969pc}.  
 
We start by reviewing the construction of the most general asymptotically flat solution of 5-d Einstein-Maxwell theory \cite{CCLP} 
with three Killing symmetries and its lift to six-dimensions.  In view of the fuzz-ball proposal \cite{Lunin:2001jy,Lunin:2002iz,Mathur:2003hj,Lunin:2004uu,Mathur:2005zp,Skenderis:2008qn,Mathur:2008nj}, according to which the micro-states of BHs should be represented  by smooth horizonless geometries, we focus on BH geometries or singular metrics that lift to a regular six-dimensional JMaRT geometry \cite{CCLP}.   We consider the supersymmetric and non-supersymmetric cases as well as extremal and non-extremal cases, and analyze the existence of a photon-sphere by employing the Hamiltonian formalism as discussed in \cite{Bianchi:2018kzy,Bianchi:2020des,Bianchi:2017sds}. 
We  stress that JMaRT geometries are to be considered as `fake' fuzz-balls in that there is no parameters that can be tuned to such a value that the (regular) solution becomes a (singular) BH with the same charges (even with different angular momenta).  
 
The plan of the paper is as follows in section \ref{section5d} we review the CCLP family of solutions of Einstein-Maxwell theory in five dimensions. In section \ref{section6d}, we discuss the lift to minimal six-dimensional supergravity. In particular, we determine the  conditions under which the metric lifts to a smooth horizon-less JMaRT geometry \cite{JMaRT}. In section \ref{sectiongeodesics} we study geodesics
near the photon-spheres of the general geometry and compute the corresponding Lyapunov exponents. We show that they always satisfy the bound $\lambda \le \lambda_{\rm Schw}$ if the singularity is hidden behind a horizon. In appendix (\ref{sectionJMaRT}) we present the dictionary between CCLP and JMaRT descriptions.

    
\section{Solutions of Einstein-Maxwell theory in five dimensions}
\label{section5d}
In this section, we describe the general  solution of Einstein-Maxwell theory with three Killing symmetries in five dimensions and Lagrangian  
\be
16\pi G_5 \,{\cal L}_{5d} =\sqrt{g_{5d}} \left (R_{5d}
 -{1\over 4} F^2 \right)-{1\over 3\sqrt{3}} F\wedge F \wedge A
\ee 
 We set  units such that the five dimensional Newton constant\footnote{The constant $G_5$ has been chosen so that the expansion at large distance of the $g_{tt}$ component of the metric reads  $g_{tt} \sim -1 + \frac{2\cal M}{r^2}$.} is $G_{5} = 3\pi/4$.  The general solution obtained in \cite{CCLP} is specified by four parameters, the mass ${\cal M}$, the charge ${\cal Q}$ and two angular momentum parameters  ${\ell_1}$, ${\ell_2}$ related to the Komar integrals  
\begin{equation}
\begin{aligned}
{\cal M} & = \frac{1}{8 \pi^2} 
\int_{S^3} * dK^{(t)} \\
Q  &= \frac{1}{4 \pi^2}  
\int_{S^3}  \left(*F- \frac{1}{\sqrt{3}} F\wedge A \right) = -\sqrt{3} {\cal Q} \\
J_{\phi} &= - \frac{1}{12 \pi^2} 
\int_{S^3} * dK^{(\phi)}   = \frac{2 {\cal M}\,\ell_1 + {\cal Q} \,\ell_2}{3} \\
J_{\psi} &= - \frac{1}{12 \pi^2} 
\int_{S^3} * dK^{(\psi)} = \frac{2 {\cal M}\,\ell_2 + {\cal Q} \,\ell_1}{3}
\end{aligned}
\end{equation}
where $K^{(i)}= g_{\mu i}  dx^\mu$, $i=t$, $\phi$, $\psi$ are the 1-forms associated to the Killing vectors $\partial_t$, $\partial_\phi$ and $\partial_\psi$.

\subsection{The metric and gauge field }

The general CCLP solution is described by the metric and gauge field \cite{CCLP}
\begin{equation}
\label{CCLPmetric}
\begin{aligned}
ds_{5}^2 &= -  dt^2  + \Delta_t \left( dt-\omega_1 \right)^2  - \frac{2\, {\cal Q} \,     }{\rho^2}  \left( dt -\omega_1 \right)  \, \omega_2+
\rho^2\, \left(   d\theta^2 +\frac{ dr^2}{\Delta_r} \right)
\\
&+ \left(r^2+\ell_1^2\right) \sin^2\theta\,d\phi^2  + \left(r^2+\ell_2^2   \right) \cos^2\theta\,d\psi^2  \\
A &= {\sqrt{3}{\cal Q} \over\rho^2} (dt-\omega_1) 
\end{aligned}
\end{equation}
where 
\begin{equation}
\begin{aligned}
\rho^2 &= r^2 +\ell_1^2\cos^2\theta + \ell_2^2\sin^2\theta
\\
\Delta_r &= \frac{\left(r^2+\ell_1^2\right) \left(r^2+\ell_2^2\right)-2{\cal M} r^2 +2 {\ell_1} {\ell_2} {\cal Q}+{\cal Q}^2}{r^2}
\quad , \quad 
\Delta_t  ={2 {\cal M} \over \rho^2}  -{{\cal Q}^2\over \rho^4}  \\
\omega_1 &= {\ell_1}\sin^2\theta\, d\phi  + {\ell_2}\cos^2\theta\, d\psi 
\quad
,
\quad \omega_2 = {\ell_2}\sin^2\theta \,d\phi  + {\ell_1}\cos^2\theta\, d\psi\end{aligned}
\end{equation}
\begin{figure}[t]
\centering
\includegraphics[scale=0.7]{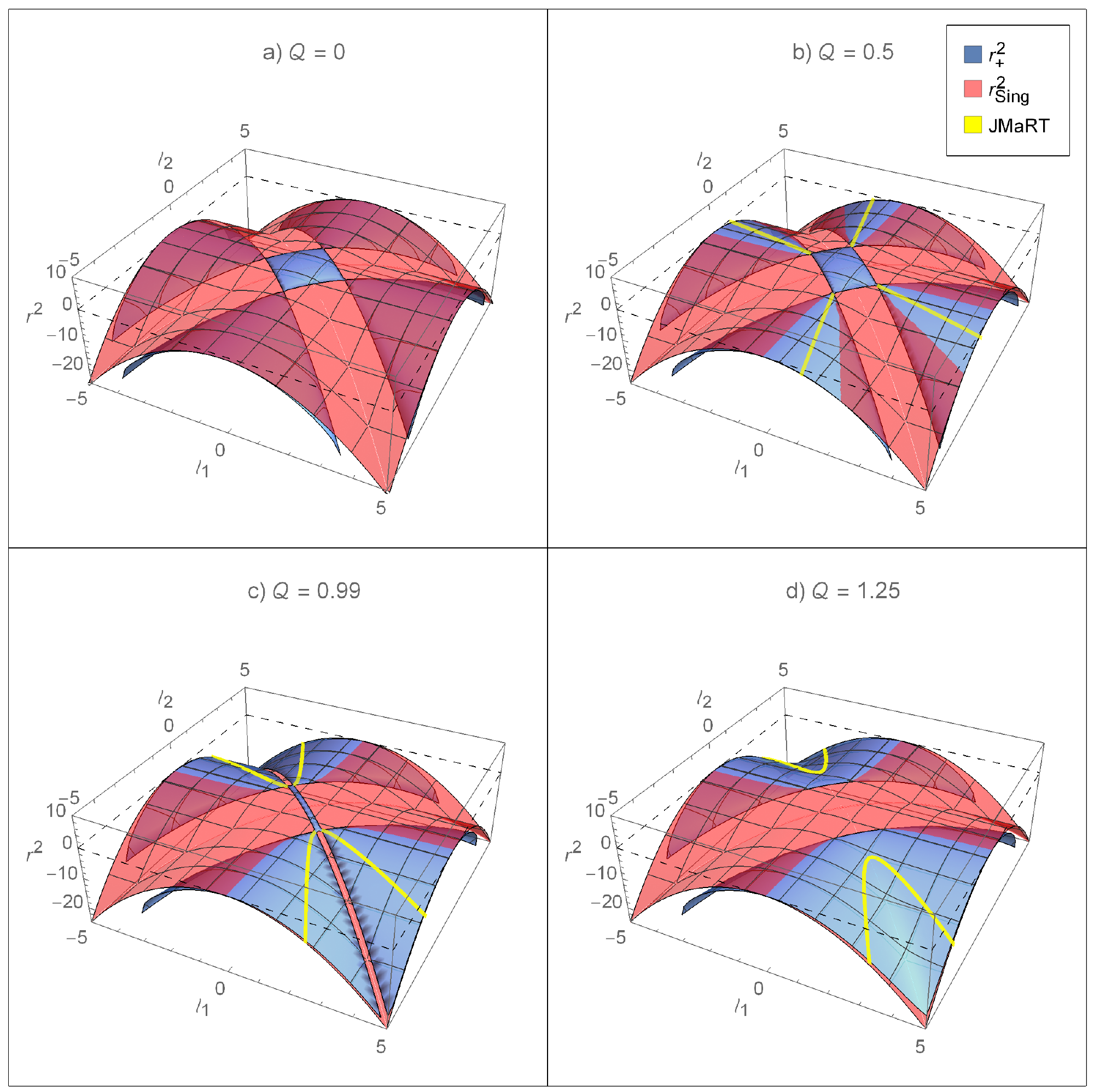}
\caption{\small{Moduli space of 5d solutions: BH's (black), regular solutions (yellow), naked singularities (red) and over-rotating geometries (blue). The mass has been set to ${\cal M} =1$, while the charge has been set to ${\cal Q} = 0,0.5,0.99,1.5$ in pictures a) to d) respectively.}}
\label{plot-singVShor}
\end{figure}

The metric has a curvature singularity at $\rho=0$,  i.e.
\begin{equation}
r_{\rm sing}^2(\theta) = -\ell_1^2\cos^2\theta - \ell_2^2\sin^2\theta
\end{equation}
and an inner and an outer horizons located at the zeros of $\Delta_r$
  \be\label{rpm}
  r_\pm^2 =\ft12   \left( 2{\cal M}- \ell_1^2-\ell_2^2  \right) \pm \ft12 \sqrt{   \left( 2{\cal M}- \ell_1^2-\ell_2^2  \right)^2 -4( {\cal Q}+\ell_1 \ell_2)^2 }
  \ee
  Finally an ergo-surface is located at the (larger) solution of $g_{tt}(r_{\rm ergo})=0$ where $\Delta_t=1$ and $\rho^2_{\rm ergo} = {\cal M} \pm \sqrt{{\cal M}^2-{\cal Q}^2}$:    
 \be
 r^2_{\rm ergo} = {\cal M} \pm \sqrt{{\cal M}^2-{\cal Q}^2}-\ell_1^2\cos^2\theta - \ell_2^2\sin^2\theta
 \ee
  
  The locations of the horizons and the curvature singularities are displayed in figure \ref{plot-singVShor}  for various choices of the charge ${\cal Q}$ (we set ${\cal M}=1$).  We display in red the maximum values of $r_{\rm sing}^2(\theta)$ for a given $\ell_1$, $\ell_2$ and in blue the location of the outer horizon $r_+$. In the $(\ell_1,\ell_2)$ plane we distinguish three different regions. The blue rectangle in the middle represents BH's where the singularity is hidden behind an event horizon. They exist for ${\cal Q}\le{\cal M}$. Red regions represent  geometries where the singularity (or a portion of it) is exposed (naked), {\it i.e.} $r^2_{\rm sing}(\theta)>r^2_+$ for some $\theta$.  Solutions  along the blue wedge/wing-like  surfaces outside the middle rectangle represents over-rotating geometries where  $r_{\rm sing}^2(\theta)<r_+^2<0$. These regions exist for $Q \neq 0$. Mind that both the singularity and the horizon, outside the central BH region, are generally hidden behind a smooth cap characterized by the vanishing of a space-like (rather than time-like) Killing vector\footnote{The near horizon geometry is better described in the $\rho$ variables.}. For the special BPS case ${\cal Q} = {\cal M}$ (BMPV solution) they have been thoroughly explored in \cite{Gibbons_1999}. It is not in the interest of this work to study the geodesic structure of the more general CCLP geometries, which has been tackled in \cite{Reimers:2016czc,Amir:2017slq}.  Finally, yellow lines host solutions that can be lifted to smooth horizon-less  six-dimensional JMaRT geometries.
  
   In the BH region, where $r_+$ given by (\ref{rpm}) is real, the BH temperature $T$ and entropy $S$ are given by
\bea
 2\pi T &=& {r_+(r_+^2 - r_-^2) \over (r_+^2+ \ell_1^2 )  (r_+^2+ \ell_2^2 )+{\cal Q} {\ell_1} {\ell_2}}\nn\\
    S &=& { 2\pi \over 3 r_+} \left[  (r_+^2+ \ell_1^2 )  (r_+^2+ \ell_2^2 )+   {\cal Q} {\ell_1} {\ell_2} \right] \label{CCLPBHentro}
  \eea

\subsection{The extremal cases}

A solutions is said to be extremal if the two horizons collide $r_+=r_-$ (so that $T=0$) and supersymmetric if ${\cal Q}={\cal M}$. 
  In this section we will discuss the two cases in turn.

 Extremal geometries are found when the argument of the square root in (\ref{rpm})
 vanishes   
 \be
{\rm Extremal:}\qquad \left| 2{\cal M}- \ell_1^2-\ell_2^2  \right| = \left| \mathcal{Q} +\ell_1 \ell_2  \right| 
 \ee
   The geometry 
  further simplifies if we require supersymmetry 
  \begin{equation}
{\rm Extremal~susy~(BPS):}\qquad {\cal Q}=\pm {\cal M} \qquad , \qquad  \ell_1 = \mp \ell_2 = \ell
\end{equation} 
  In this limit the angular momenta satisfy $J_\phi=\mp J_\psi = {\cal M}\ell/3$. The metric takes the simple form  \cite{BMPV}
  \be\label{BMPVbh}
ds^2_{BMPV} = - Z^2 \left[dt + {\ell\,{\cal M}\over \rho^2 Z } (\sin^2\theta \,d\phi - \cos^2\theta\, d\psi)\right]^2+ {d\rho^2 \over Z^2} + \rho^2ds^2_{S^3}
\ee
 with $\rho^2 = r^2 + \ell^2$ and
\be
Z = 1 - {{\cal M}\over \rho^2} \qquad , \qquad
ds^2_{S^3}=d\theta^2 + \sin^2\theta \,d\phi^2+ \cos^2\theta\, d\psi^2\,,
\ee
   The solution has a singularity at $\rho=0$, no ergo-region and a horizon at $\rho^2={\cal M}$. For $\ell^2<{\cal M}$, this is an event horizon and
    the BH is known as the BMPV BH after Breckenridge, Myers, Peet and Vafa \cite{BMPV}.

\section{The six-dimensional uplift} 
\label{section6d}

The five-dimensional solution we described in the last section can be embedded in string theory and ten-dimensional supergravity.
In string theory, `large' BPS BH's in five-dimensions require bound-states of strings and branes with at least 3 charges. The most studied case is the bound-state in Type IIB superstring of $N_1$ D-strings (D1-branes), wrapped on a circle $S^1$, 
$N_5$ D-pentabranes (D5-branes), wrapped on a 5-torus $S^1\times T^4$, and KK momentum $N_p$ along $S^1$. The system can be described as a solution of gravity theory in five-dimensions coupled to 3 abelian photons with charges $Q_1$, $Q_5$, $Q_p$ and scalar fields.  The ten-dimensional metric can be written (in the string frame) as  
\be
ds^2_{10} = {1\over\sqrt{H_1H_5}} [-dt^2{+}dy^2{+}(H_p{-}1)(dt-dy)^2] +
\sqrt{H_1H_5} d\vec{x}d\vec{x} + \sqrt{H_1\over H_5} d\vec{z}d\vec{z}
\ee
where $\hat{r}^2 = \vec{x}^2$, $H_i = 1 + {Q_i\over \hat{r}^2}$, 
 \be
Q_1 = N_1 {g_s {\alpha'}^3 \over V_4} \quad , \quad 
Q_5 = N_5 {g_s {\alpha'}} \quad , \quad 
Q_p = N_p {g_s^2 {\alpha'}^4 \over R^2 V_4}
\ee
$R$ being the radius of $S^1$ and $(2\pi)^4V_4$ the volume of $T^4$. The metric is supported by a non-trivial profiles for the dilaton, the NSNS and RR two-forms.  
After reduction to five dimensions, one finds a three charge BPS BH  whose horizon is an $S^3$ sphere of area  
\be
A_H = 2\pi^2 \sqrt{Q_1Q_5Q_p} = 8 \pi G_5\,\sqrt{N_1N_5N_p} = 4 G_5\,S_{BH}
\ee
and $G_5= 8 \pi^6 g_s^2 {\alpha'}^4/(2\pi)^5 R V_4$. Alternatively, the D1-D5-p solution can be obtained starting from a neutral   
Schwarzschild BH in five dimensions after a lift to ten dimensions and a sequence of boosts, S and T dualities \cite{Giusto:2004id}.
Similarly, a rotating version of this system can be obtained starting from a five-dimensional rotating BH \cite{Cvetic:1998xh}.

The solution drastically simplifies if one takes all charges to be equal
$Q_1=Q_5=Q_p$. For this choice the dilaton and the metric of $T^4$ become constant and we can view the solution as a
solution of minimal supergravity in six-dimensions involving just the metric and a two-form with anti-self dual field strength. The resulting background provides a six-dimensional lift of the CCLP solution discussed in the previous section and will be the main subject of 
our discussion in this section.

\subsection{Solutions of minimal six-dimensional supergravity}

As anticipated above, solutions of Einstein-Maxwell theory can be lifted to  solutions of six-dimensional gravity coupled to an anti-self dual two-form field $C_2$, with Lagrangian\footnote{Very much as in $D=10$, the anti-self-duality constraint does not follow from the Euler-Lagrange equations but should be imposed by hand.} \cite{Giusto:2004id, Cveti__1996}
\begin{equation}
16 \pi\,G_6 \,\mathcal{L}_{6d} =  \sqrt{-g_{6d}}\left(R_{6d} - \frac{1}{12}H_3^2\right)\,,
\qquad
H_3 = d C_2 = -*H_3
\end{equation}
where $G_6=2 \pi R~ G_5$. 
The metric and two-form are given by
\begin{equation}
\label{metric6d}
\begin{aligned} 
ds_6^2 &=(dy-\frac{A}{\sqrt{3}})^2 +ds^2_5
\\
C_2 &= -\frac{\mathcal{Q}}{\rho^2}\left[\cos^2\theta\left(\ell_1 dt + \ell_2 dy\right)\wedge d\psi
+
\sin^2\theta\left(\ell_2 dt + \ell_1 dy\right)\wedge d\phi
+\right.
\\
&\left.+
dt\wedge dy + \left(r^2 + \ell_1^2\right)\cos^2\theta d\psi\wedge d\phi\right]
\end{aligned}
\end{equation}
with $ds_5^2$ and $A$ the five-dimensional metric and gauge field, respectively, given in (\ref{CCLPmetric}).  Alternatively, the same solution (\ref{metric6d}) is found starting from the non-supersymmetric D1-D5-p background in \cite{Giusto:2004id} by taking all brane charges to be equal (see Appendix \ref{sectionJMaRT} for details).  

\subsection{JMaRT geometries}

In this section we determine the conditions under which the metric (\ref{metric6d}) is regular everywhere. The resulting geometries  belong to the well known family of JMaRT solutions \cite{JMaRT},  and are obtained from the general case by setting all charges to be equal. The precise dictionary between CCLP and JMaRT variables is worked out in the Appendix. 

 A regular six-dimensional  geometry is characterised by the existence of  a Killing vector
 \be
  \partial_\xi = \partial_y- \alpha \partial_\psi-\beta \partial_\phi
\ee
vanishing at a point where the four-dimensional space spanned by $(y,\theta,\phi,\psi)$ shrinks to zero size. 
   These conditions ensure that the space terminates at cap with $\mathbb{R}^2$ topology in the $(r,\xi)$ plane. The cap can be conveniently located at the outer horizon $r_+$. To ensure that the curvature singularity is outside the space we require that $r_+^2 >r_{\rm sing}^2$, {\it i.e.}
  \begin{equation}
r_+^2 > -\ell_{1,2}^2.  \label{rpbound2}
\end{equation}
  A six-dimensional geometry is regular if the following conditions hold
  \be
{\rm det} \, g_{_{y\theta\phi\psi}}|_{r=r_+} =\Delta(r_+)=  \partial_\xi  \cdot  \partial_\xi  \big |_{r=r_+} =0
  \ee
  for any $\theta$, where ${\rm det} \, g_{_{y\theta \phi\psi}}$ is the determinant of the metric in the subspace with $t=\text{const}$ and $r=r_+$. It can be checked that the last equation gives two independent conditions, so the regularity conditions can be solved
  for ${\cal M}$, ${\cal Q}$, $\alpha$, $\beta$ in terms of $r_+$, $\ell_1$, $\ell_2$\footnote{In terms of the angular momenta one has
  $\alpha = {\cal Q}/(3 J_\psi)$, $\beta = {\cal Q}/(3 J_\phi)$.}
 \begin{equation}
\begin{aligned}\label{mqJMaRT}
\mathcal{M} &= \frac{\left(r_+^2+\ell _1^2\right) \left(r_+^2+\ell _2^2\right) \left(r_+^2+\ell _1^2+\ell _2^2\right)}{2 \ell _1^2 \ell _2^2}\\
{\cal Q} &= -\frac{\left(r_+^2+\ell _1^2\right) \left(r_+^2+\ell _2^2\right)}{\ell_1 \ell_2} \\
\alpha &  = -\frac{\ell _1}{r_+^2+\ell _2^2}\,,
\qquad
\beta   = -\frac{\ell _2}{r_+^2+\ell _1^2}
\end{aligned}
\end{equation}
We notice that the condition (\ref{rpbound2}) ensures that the mass is positive and that ${\cal Q}\, \ell_1 \,\ell_2<0$. Moreover, since the area of the horizon vanishes, the asymptotics of the JMaRT geometry cannot be considered as a BH microstate geometry.  
  
Finally, one has to impose the absence of orbifold singularities at $r=r_+$. To this end, it is convenient to introduce 
 the new (tilded) variables
\begin{equation}\label{angular}
r = \sqrt{\tilde{r}^2 + r_+^2}\,,
\qquad
\phi = \tilde{\phi} - \beta y\,,
\qquad
\psi = \tilde{\psi} - \alpha y\,,
\end{equation}
such that the cap is located at $\tilde{r} =0$ and the vanishing Killing vector is $ \partial_y$. 
Plugging (\ref{angular}) into the six-dimensional metric and expanding around $\tilde{r}\approx 0$ one finds that the metric near the origin of  the two-dimensional 
$(\tilde r,y)$-space  reduces to the flat metric
\begin{equation}
\begin{aligned}\label{ds2d}
ds^2 &\approx \rho^2 \left[\frac{\tilde{r}^2 \left[ \left(r_+^2+\ell _1^2+\ell _2^2\right)^2-\ell _1^2 \ell _2^2\right]dy^2}{\left(r_+^2+\ell _1^2\right){}^2 \left(r_+^2+\ell _2^2\right){}^2} 
-
\frac{\ell _1^2 \ell _2^2 d\tilde{r}^2}{r_+^2\left[\left(r_+^2+\ell _1^2+\ell _2^2\right)^2-\ell _1^2 \ell _2^2\right]}\right]
\end{aligned}
\end{equation}
The metric is positive definite if $r_+^2$ falls within the range
\be
| \ell_1 \ell_2 | -\ell_1^2-\ell_2^2< r_+^2<0  
\ee
 To avoid orbifold singularities at the origin in (\ref{ds2d}), we require that  the $y$ coordinate is compactified on a circle $y\sim y+2\pi R$ with radius
\begin{equation}
\label{y-circle_radius}
R = \frac{1}{\sqrt{-r_+^2}  }\frac{\left(r_+^2+\ell _1^2 \right)^2\left(r_+^2+\ell _2^2 \right)^2\left|\ell _1 \ell _2\right|}{\left(r_+^2+\ell_1^2+\ell_2^2 \right)^2-\ell_1^2\ell_2^2}\,,
\end{equation}
In addition the angular identifications (\ref{angular}) require the quantization conditions
\begin{equation}
\begin{aligned}
\label{quantization_rules}
\alpha R & = \frac{1}{\sqrt{-r_+^2} }\frac{\left(r_+^2+\ell _1^2 \right)^2\ell_1\left|\ell _1 \ell _2\right|}{\left(r_+^2+\ell_1^2+\ell_2^2 \right)^2-\ell_1^2\ell_2^2} = -n\\
\beta R &= \frac{1}{\sqrt{-r_+^2} }\frac{\left(r_+^2+\ell _2^2 \right)^2\ell_2\left|\ell _1 \ell _2\right|}{\left(r_+^2+\ell_1^2+\ell_2^2 \right)^2-\ell_1^2\ell_2^2} =  m
\end{aligned}
\end{equation}
These equations can be solved for $\ell_1$, $\ell_2$ in terms of $m$, $n$. For example, assuming $m>n>0$ positive one finds  
\begin{equation}
\label{ell1_ell2_eqs}
\begin{aligned}
\ell _1 &=  { \sqrt{-r_+^2} \over m   }  \left(\nu_{m,n}  +{m^2 n^2 \over \nu_{m,n} }+ m^2 \right)
\\
\ell _2 &= - { \sqrt{-r_+^2} \over n   }  \left(\nu_{m,n}  +{m^2 n^2 \over \nu_{m,n} }+ n^2 \right)
\end{aligned}
\end{equation}
with
\begin{equation}\label{nu}
\nu_{m,n}^3 = { m^2 n^2 \over 2}  \left( m^2 +n^2-1- \sqrt{ ( m^2 -n^2-1)^2  -4 n^2 }     \right) 
\end{equation}

\subsection{Supersymmetric solution}

The supersymmetric limit of the JMaRT solution, known as  GMS  solution (after Giusto, Mathur and Saxena) \cite{Giusto:2004id},  is obtained by taking $m = n+1$. For this choice, the argument of the square root in (\ref{nu}) vanishes and one finds 
\begin{equation}
\label{ell1_ell2_GMS_eqs}
{\cal M}={\cal Q}=-9n(n+1) r_+^2 \qquad \ell _1 = \sqrt{-r_+^2}(3 n+ 1)\,,
\qquad
\ell _2 = -\sqrt{-r_+^2} (3 n + 2)
\end{equation}
  For large $n$, this solution can be viewed as the fuzzball of a BPMV BH with 
  \be
  {\cal M}={\cal Q}=\ell_1^2=\ell_2^2 =-9 n^2 r_+^2
  \ee

\section{Critical geodesics in the 5d geometries}
\label{sectiongeodesics}
After this long preamble on five-dimensional BHs and  their regular 6d uplifts  (`fake' fuzzballs), we are ready to tackle the main issue 
of the present work: the study of geodesics of neutral massless particles moving in the vicinity of the light rings of  five-dimensional CCLP geometries. We will show that geometries will always exhibit a photon-spheres surrounding the mass center and compute
  the Lyapunov exponents $\lambda$ characterizing the chaotic behaviour of geodesics near the light rings.  
  We show that for  any acceptable solution -- in view of the cosmic censorship hypothesis -- including BHs, over-rotating solutions and solutions admitting a 
regular horizon-free six-dimensional lifts, $\lambda$ is always bounded by its value for a Schwarzschild BH of the same mass. The bound will be however violated at the internal photon-sphere of geometries exposing a naked singularity where the light ring crashes against the singularity.  

Null geodesics are simply solutions to the equation  \be
{ds^2_5\over  d\tau^2}=g_{\mu\nu} \dot{x}^\mu \dot{x}^\nu = 0\ee where $\dot{x}^\mu$ are the generalised velocities and $\tau$ is some affine parameter describing the geodesic. For metrics with a large group of isometries it is convenient to switch to the Hamiltonian formalism whereby the dynamics of a massless neutral particle is governed by the equivalent  condition
\be
\mathcal{H} = \frac{1}{2}\,g^{\mu\nu} P_\mu P_\nu = 0 
\ee
with $P_\mu=g_{\mu\nu} \dot{x}^\nu$ the conjugate momenta. In the cases under consideration we have three conserved momenta:  
 $P_t=-E$, $P_\phi = J_\phi$ and $P_\psi = J_\psi$. Moreover the Hamiltonian can be separated as
\begin{equation}
\mathcal{H} = \frac{\Lambda_r + \Lambda_\theta}{2 \rho^2} = 0
\end{equation}
with  
\begin{equation}
\label{sepCCLP}
\small
\begin{aligned}
&\Lambda_r = \frac{E^2}{r^2 \Delta _r}\left[\left(r^2+\ell _1^2+\ell _2^2\right) \left({\cal Q}^2-r^2 \Delta _r\right)-2{\cal M} \left(r^2+\ell _1^2\right) \left(r^2+\ell _2^2\right)\right]+P_r^2 \Delta _r 
+
\\
&+
\frac{P_{\psi }^2 }{r^2 \Delta _r}\left[\left(\ell _1^2-\ell _2^2\right) \left(r^2+\ell _1^2\right)-2 \ell _1 \left(\mathcal{M}\,\ell _1{\,+\,}{\cal Q} \ell _2\right)\right]
+\frac{P_{\phi }^2 }{r^2 \Delta _r} \left[\left(\ell _2^2-\ell _1^2\right) \left(r^2+\ell _2^2\right)-2 \ell _2 \left(\mathcal{M}\,\ell _2{\,+\,}{\cal Q} \ell _1\right)\right]
+
\\
&+
\frac{2 P_{\psi }\,E }{r^2 \Delta _r}
\left[\left(r^2+\ell _1^2\right) \left(2{\cal M}\,\ell _2+{\cal Q} \ell _1\right)-\ell _2 \left({\cal Q}^2+r^2 \Delta _r\right)\right]
-\frac{2 P_{\psi } P_{\phi } }{r^2 \Delta _r}
\left[2{\cal M}\,\ell _1 \ell _2+{\cal Q} \left(\ell _1^2+\ell _2^2\right)\right]
+
\\
&+
\frac{2 P_{\phi }\,E }{r^2 \Delta _r}\left[\left(r^2+\ell _2^2\right) \left(2{\cal M}\,\ell _1+{\cal Q} \ell _2\right)-\ell _1 \left({\cal Q}^2+r^2 \Delta _r\right)\right]= -K^2
\\
&\Lambda _{\theta } = P_{\theta }^2+\left(E\,\ell_1\sin\theta +\frac{P_{\phi }}{\sin\theta}\right)^2+\left(E\,\ell_2\cos\theta +\frac{P_{\psi }}{\cos\theta}\right)^2 = K^2
\end{aligned}
\end{equation} 
As apparent, the separation constant $K^2$ plays the role of the square of the total angular momentum.
The radial velocity of a massless particle falling into the BH geometry is
\begin{equation}
\frac{dr}{dt} =-{  \frac{\partial \cal H }{\partial P_r}\over 
\frac{\partial \cal  H }{\partial E} } = \frac{\sqrt{\mathcal{R}(r)}}{r\,\mathcal{S}(r,\theta)}
\end{equation}
where
\begin{align}
\small
&\mathcal{R}(r) = \left(\frac{r}{2E}\frac{\partial \Lambda_r}{\partial P_r}\right)^2 = \frac{r^2\Delta_r^2 P_r^2}{4E^2} = A r^6 + B r^4 + C r^2 + D\\
&\mathcal{S}(r,\theta) = r^2 +\ell_1^2\cos^2\theta + \ell_2^2\sin^2\theta + \frac{1}{r^2\Delta_r}\left[b_{\phi } \left({\cal Q}^2 \ell _1-{\cal Q} \ell _2 \left(r^2+\ell _2^2\right)-2{\cal M}\,\ell _1 \left(r^2+\ell _2^2\right)\right)\right.
+  \label{non-extRandS}
\\
&+
\left.b_{\psi } \left({\cal Q}^2 \ell _2{-}{\cal Q} \ell _1 (r^2{+}\ell _1^2){-} 2{\cal M}\,\ell_2 (r^2{+}\ell _1^2)\right) + 2{\cal M} (r^2{+}\ell_1^2) (r^2{+}\ell _2^2)-{\cal Q}^2 (r^2{+}\ell _1^2{+}\ell_2^2)\right] \nn
\end{align}
with
\begin{equation}
\small
\begin{aligned}
A &=1
\quad
,
\quad
B=2 \ell _2 b_{\psi }+2 \ell _1 b_{\phi }+2 \left(\ell _1^2+\ell _2^2\right)-b^2
\quad
,
\quad\\
C & = 2 {\cal Q} \ell _2 \ell _1+\ell _1^4+3 \ell _2^2 \ell _1^2+\ell _2^4 + b^2 \left(2{\cal M}-\ell _1^2-\ell _2^2\right)+\left(\ell _1^2-\ell _2^2\right)\left( b_{\phi }^2 - b_{\psi }^2\right)
+
\\
&+
2b_{\phi } \left(\ell _1 \left(\ell _1^2+\ell _2^2 -4 \mathcal{M}\right) - {\cal Q} \ell _2\right)
+
2b_{\psi } \left(\ell _2 \left(\ell _1^2+\ell _2^2 -4 \mathcal{M}\right) - {\cal Q} \ell _1\right)
\\
D &=\ell _1 b_{\psi }^2 \left(2{\cal M} \ell _1+2 {\cal Q} \ell _2-\ell _1^3+\ell _2^2 \ell _1\right)+\ell _2 b_{\phi }^2 \left(\ell _2 \left(2{\cal M}+\ell _1^2-\ell _2^2\right)+2 {\cal Q} \ell _1\right)
+
\\
&+
\ell _1 \ell _2 \left(\ell _1 \ell _2 \left(2{\cal M}+\ell _1^2+\ell _2^2\right)+2 {\cal Q} \left(\ell _1^2+\ell _2^2\right)\right) -b^2 \left({\cal Q}+\ell _1 \ell _2\right)^2
+
\\
&+
2b_{\phi}b_{\psi } \left(2{\cal M} \ell _1 \ell _2+ {\cal Q} \left(\ell _1^2+\ell _2^2\right)\right)
+
b_{\psi } \left(2 \ell _1^2 \ell _2 \left(\ell _2^2-2{\cal M}\right)+4 {\cal Q}^2 \ell _2-2 {\cal Q} \left(\ell _1^3-2 \ell _1 \ell _2^2\right)\right) 
+
\\
&+
b_{\phi } \left(2 \ell _2^2 \ell _1 \left(\ell _1^2-2{\cal M}\right)+4 {\cal Q}^2 \ell _1-2 {\cal Q} \left(\ell _2^3-2 \ell _2 \ell _1^2\right)\right)
\end{aligned}
\end{equation}
and
\begin{equation}
b = \frac{K}{E}
\quad
,
\quad
b_\phi = \frac{P_\phi}{E}
\quad
,\quad
b_\psi = \frac{P_\psi}{E}
\end{equation}
are the impact parameters.

For simplicity, we focus on geodetic motion with constant $\theta=\theta_0$. This is only allowed  on the hyper-planes $\theta_0=0,\pi/2$. More precisely,
the geodesics along these hyper-planes are characterised by the following impact parameters
\be
\label{constant_theta_motion}
\begin{aligned}
\theta_0 &= 0 \qquad , \qquad &b_\phi = 0 \qquad , \qquad b &= b_\psi + \ell_2  \\
\theta_0 &= {\pi/2} \qquad , \qquad &b_\psi = 0 \qquad , \qquad b &= b_\phi +\ell_1   
\end{aligned}
\ee
 For definiteness we will further specialize to the plane $\theta_0 = 0$. The results for $\theta=\pi/2$ can be obtained from the ones presented here by exchanging $\ell_1 \leftrightarrow \ell_2$. 
 
We are interested in the chaotic behaviour of critical geodesics. A geodesics is said to be critical if it reaches the turning point with zero radial acceleration. This happens for choices of the impact parameters such that the critical conditions
\be
 \mathcal{R}(r_c)=\mathcal{R}'(r_c)=0 \label{critical}
\ee
admit a solution $r_c$. At this critical radius, photons can orbit along unstable null circular orbits forming the so called photon-sphere, in our case a light-ring since $\theta=0$ is fixed.   
Near the critical radius one finds  
 \be
 {\cal R}(r) \approx {1\over 2} {\cal R}''(r_c)(r-r_c)^2 \approx 4 \, r_c^2  ( 3 \, A\, r_c^2+ B ) \,(r-r_c)^2  
 \ee
  As a result the radial velocity reduces to
 \be
 {dr\over dt}\approx  - 2 \lambda (r-r_c) \label{radial0}
 \ee
 with Lyapunov exponent
 \be
\lambda= {   \sqrt{  {\cal R}''(r_c)/2} \over  2\, r_c\, {\cal S}_0(r_c) }   = {   \sqrt{ 3 \, r_c^2+ B } \over    {\cal S}_0(r_c) }   
\label{lambda}
\ee
where ${\cal R}(r)$, ${\cal S}_0(r_c)={\cal S}(r_c,0)$ 
are given by (\ref{non-extRandS}) evaluated at $\theta=0$ with $b_\phi = 0$ and $b = b_\psi + \ell_2 $. 
For this choice one finds
\begin{equation}
\label{poly_coeff}
\small
\begin{aligned}
A &=1
\quad
,
\quad
B=2 b \ell _2+2 \ell _1^2 -b^2
\quad
,
\quad\\
C & = -2 b^2 \left(\ell _1^2-\mathcal{M}\right)-2 b \left(\mathcal{Q} \ell _1+4 \ell _2 \mathcal{M}-2 \ell _2 \ell _1^2\right)+4 \mathcal{Q} \ell _2 \ell _1+8 \ell _2^2 \mathcal{M}+\ell _1^4
\\
D &=- (b-2 \ell _2 ) \left[b \left(\mathcal{Q}^2-2 \ell _1^2 \mathcal{M}+\ell _1^4\right)-2 \mathcal{Q}^2 \ell _2+2 \mathcal{Q} \ell _1^3+4 \ell _2 \ell _1^2 \mathcal{M}\right]
\end{aligned}
\end{equation}
while $\mathcal{S}_0(x=r_c^2)$ reads
\begin{equation}
\begin{aligned}\label{s0}
\mathcal{S}_0(x) 
&=
\ell _1^2+x+\frac{2 \mathcal{M} \left(\ell _1^2+x\right) \left(\ell _2^2+x\right)-\mathcal{Q}^2 \left(\ell _1^2+\ell _2^2+x\right)}{\mathcal{Q}^2+2 \ell _1 \ell _2 \mathcal{Q}-2 x \mathcal{M}+\left(\ell _1^2+x\right) \left(\ell _2^2+x\right)}
+
\\
&+
\frac{\left(b-\ell _2\right) \left(\ell _2 \mathcal{Q}^2-\ell _1 \left(\ell _1^2+x\right) \mathcal{Q}-2 \mathcal{M} \left(\ell _1^2+x\right) \ell _2\right)}{\mathcal{Q}^2+2 \ell _1 \ell _2 \mathcal{Q}-2 x \mathcal{M}+\left(\ell _1^2+x\right) \left(\ell _2^2+x\right)}
\end{aligned}
\end{equation}
In the following we evaluate these formulae for various choices of charges and angular momenta. The mass $\mathcal{M}$ set the scale of the geometry and can always be taken to be one.

\subsection{Examples}

\subsubsection{Schwarzschild-Tangherlini BH}

To obtain the Schwarzschild-Tangherlini BH we take ${\cal Q} = \ell_1 = \ell_2 = 0$ in \ref{poly_coeff}, the polynomial is simply
\begin{equation}
\mathcal{R}(x) = -b^2 x^2+2 b^2 x \mathcal{M}+x^3\,,
\qquad
\mathcal{S}(x) = \frac{x^2}{x-2 \mathcal{M}}
\end{equation}
where $x = r^2$. The critical radius, impact parameter and Lyapunov exponent read
\begin{equation}
r_c = 2\sqrt{\cal M}
\qquad
,
\qquad
b_c = 2\sqrt{2{\cal M}}
\qquad
,
\qquad
\lambda_{\rm Schw} = \frac{1}{\sqrt{2}\,b_c}={1\over 4 \sqrt{\cal M}}
\end{equation}

\subsubsection{Reissner-Nordstr{\"o}m extremal BH}
%
Next we consider a charged non rotating BH in the extremal (BPS) limit
\be
\mathcal{Q} = \mathcal{M}\,,
\qquad
\ell_1=\ell_2=0
\ee
the polynomial ${\cal R}(r)$ and the function ${\cal S}_0(r)$  reduce to
 \begin{equation}
\begin{aligned}
 {\cal R}(r) &=r^6 - b^2( r^2 -{\cal M}  )^2\,,
 \qquad
 \mathcal{S}_0(r) &=   {r^6\over (r^2-{\cal M})^2}
\end{aligned}
\end{equation}
 The critical equations ${\cal R}'(r)={\cal R}(r)=0$ are solved by  
  \be
 r_c =\sqrt{3 {\cal M} }  \qquad ,\qquad  b_c={3\sqrt{3 {\cal M} } \over 2}
 \ee
leading to the Lyapunov exponent  
  \be
  \lambda_{RN}= \frac{1}{3b_c} =    
   {  2\over 9 \sqrt{{\cal M}} } < \lambda_{\rm Schw} 
    \ee
\subsubsection{Uncharged rotating solutions}

\begin{figure}[t]
\centering
\subfloat[]{\includegraphics[scale=0.37]{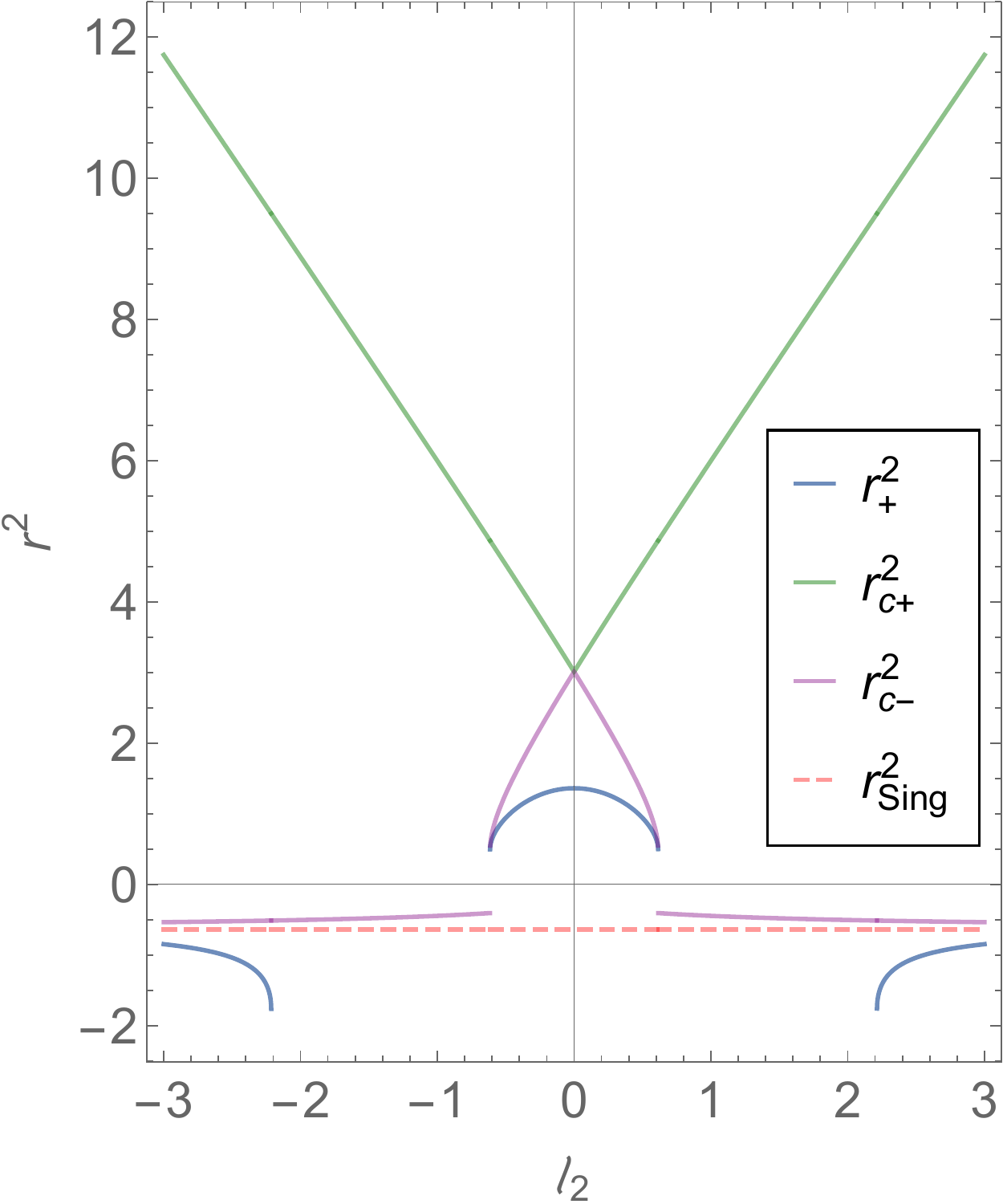}}
\quad
\subfloat[]{\includegraphics[scale=0.37]{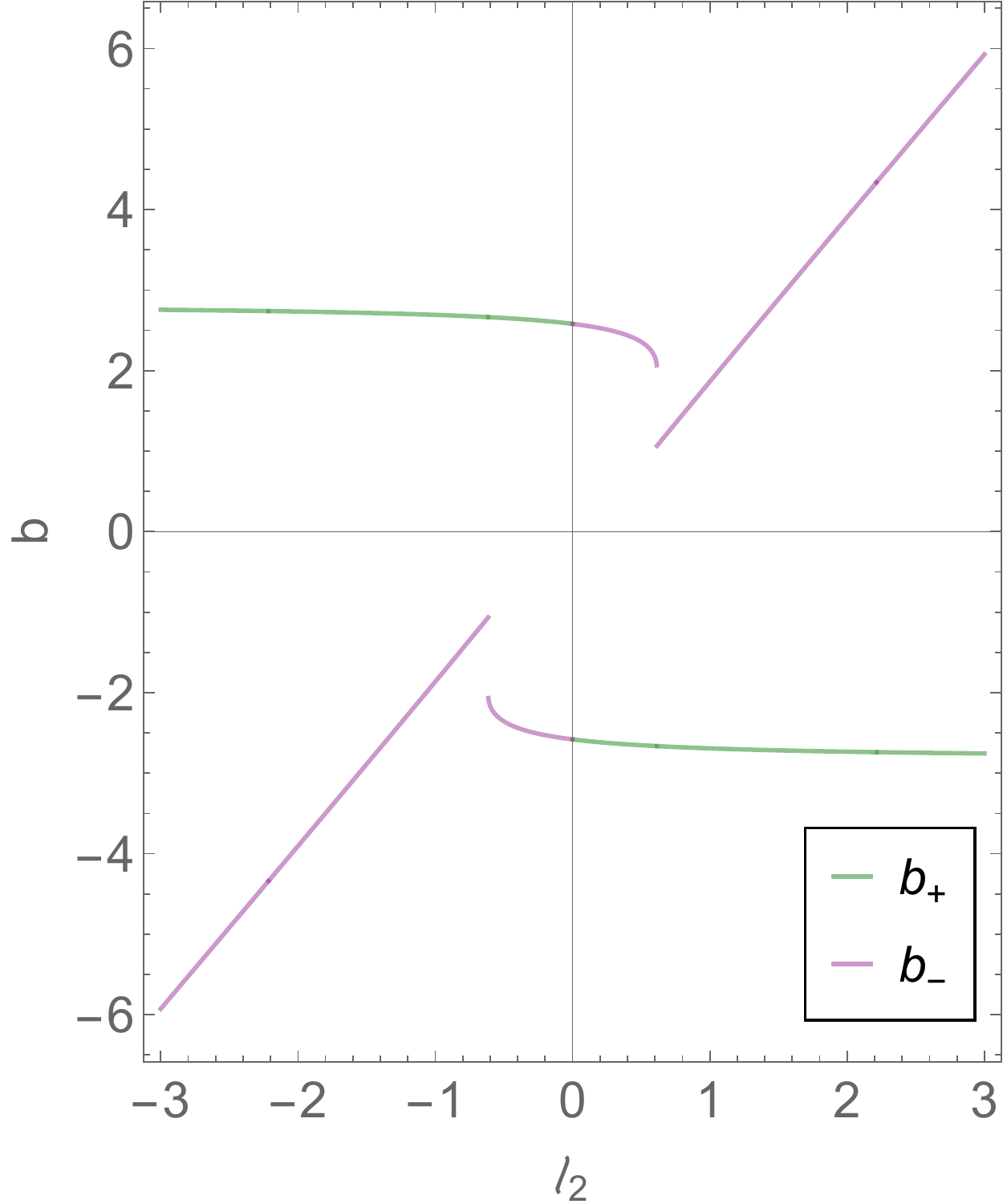}}
\quad
\subfloat[]{\includegraphics[scale=0.37]{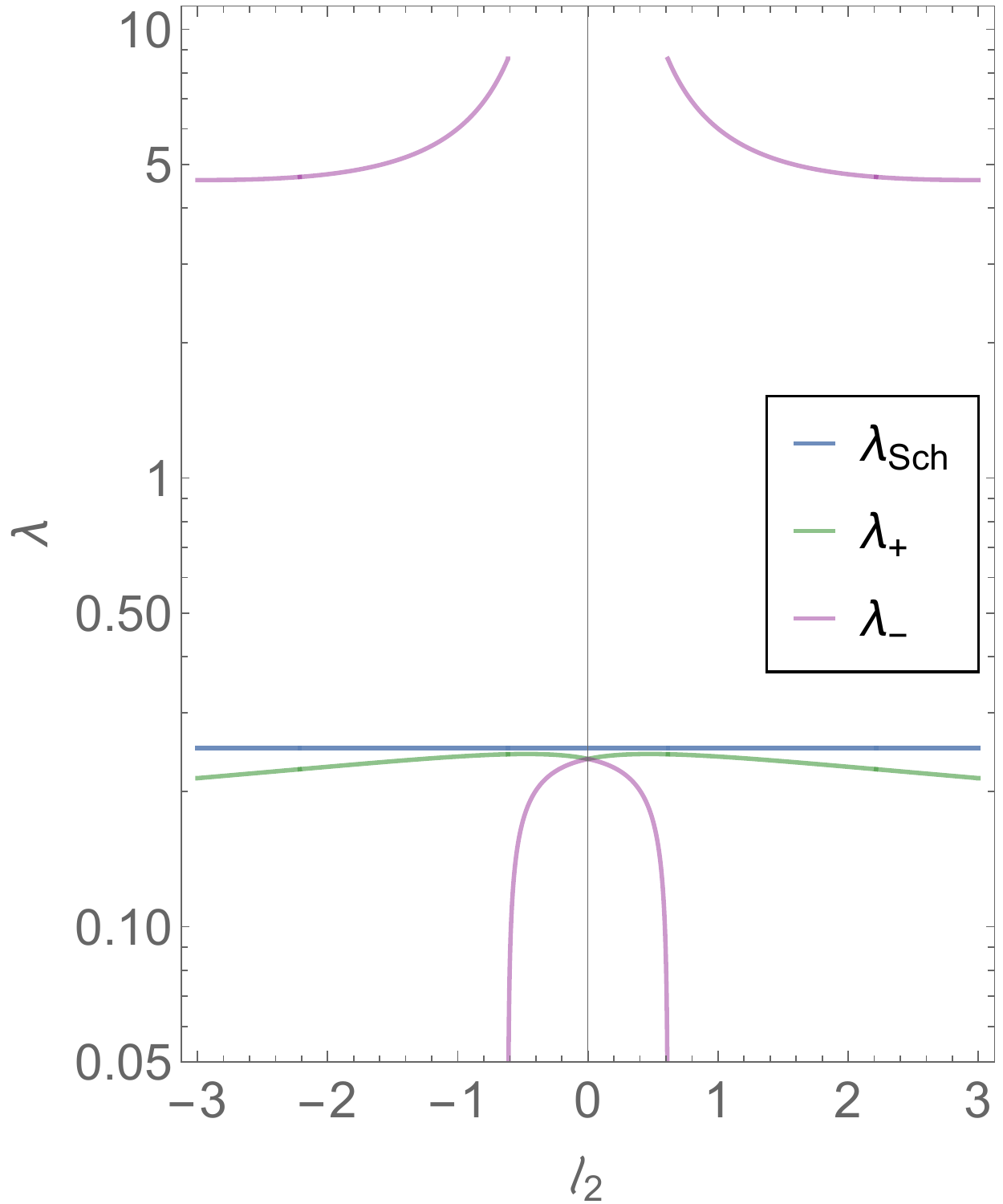}}
\caption{\small{a) In red the maximum values of $r_{\rm sing}^2$, in blue the values of $r_+^2$, In green and purple the values of $r_{c,+}^2$ and $r_{c,-}^2$ respectively. b) In green and purple the values of $b_{c,+}$ and $b_{c,-}$ respectively. c) In green and purple the values of $\lambda_{+}$ and $\lambda_{-}$ respectively. In all plots we set ${\cal M} =1$ and $\ell_1 = 0.8$.}}
\label{plot-Q0singVShor}
\end{figure}

The case ${\cal Q} = 0$ with arbitrary angular momenta (also known as  Myers-Perry BH) provides a nice illustration of the general case. For this choice the polynomial 
${\cal R}(x)$ factors into a quadratic times a linear polynomial in $x=r^2$, so analytic solutions for the critical radius can be found. 
From (\ref{non-extRandS}) one finds
\begin{equation}
\begin{aligned}
\mathcal{R}(r^2)&= \left(r^2+\ell _1^2\right) \left[\left(r^2+\ell _1^2\right) \left(-b^2+2 b \ell _2+r^2\right)+2 \mathcal{M} \left(b-2 \ell _2\right)^2\right]
\\
\mathcal{S}_0(r^2)&=\frac{\left(r^2+\ell _1^2\right) \left(2 \ell _2 \mathcal{M} \left(2 \ell _2-b\right)+\left(r^2+\ell _1^2\right) \left(r^2+\ell _2^2\right)\right)}{\left(r^2+\ell _1^2\right) \left(r^2+\ell _2^2\right)-2 r^2 \mathcal{M}}
\end{aligned}
\end{equation}
There are four choices of the impact parameter $b_c$ for which the critical conditions $\mathcal{R}(r_c)=\mathcal{R}'(r)=0$ admit a solution.
\begin{equation}
b_c = \ell _2+\sigma_1 \sqrt{2\mathcal{M}} +\sigma_2 \sqrt{\left(\ell _2-\sigma_1 \sqrt{2\mathcal{M}}\right)^2-\ell _1^2}
\end{equation}
with $\sigma_1,\sigma_2=\pm$. The critical radius can be neatly expressed in terms of the critical impact parameter
\begin{equation}
\label{Q0critsol}
\begin{aligned}
r_c^2&= \frac{1}{2} \left(b_c^2-2 b_c\,\ell _2-\ell _1^2\right)\,,
\end{aligned}
\end{equation}
where $b_c$ is chosen such that $r_c$ is a minimum. There are two choices leading to minima, denoted by $r_{c-},r_{c+}$ with
$r_{c-}<r_{c+}$, that represent the inner and outer radius of the photon-sphere. 
For the Lyapunov exponent  one finds \begin{equation}
\lambda =\frac{\left(b_c-2 \ell _2\right)^2-\ell _1^2}{\sqrt{2} \sqrt{b_c \left(b_c-2 \ell _2\right)+\ell _1^2} \left(b_c^2-3 b_c \ell _2+2 \ell _2^2\right)}
\end{equation}
 In figures \ref{plot-Q0singVShor}, we display the dependence on $\ell_2$ (we set ${\cal M}=1$, ${\cal Q}=0$, $\ell_1=0.8$) of the critical radius $r_{c\pm}$, the critical impact parameters $b_{c\pm}$ and the Lyapunov exponents $\lambda_{\pm}$ 
 for the two solutions: green (+) and purple (-). We notice that $\lambda_+$ (the exponent at the outer photon-sphere) is always below the bound $\lambda_{\rm Schw}$, while $\lambda_-$ is below the bound only inside the BH region where the singularity is hidden behind a horizon. Outside this region, the inner photon-sphere crashes against the naked singularity and the Lyapunov exponent blows up. Remarkably, the cosmic censorship hypotesis, ruling out the existence of naked singularities, prevents the violation of the bound!

\subsubsection{Extremal supersymmetric rotating solution}
\begin{figure}[t]
\centering
\subfloat[]{\includegraphics[scale=0.37]{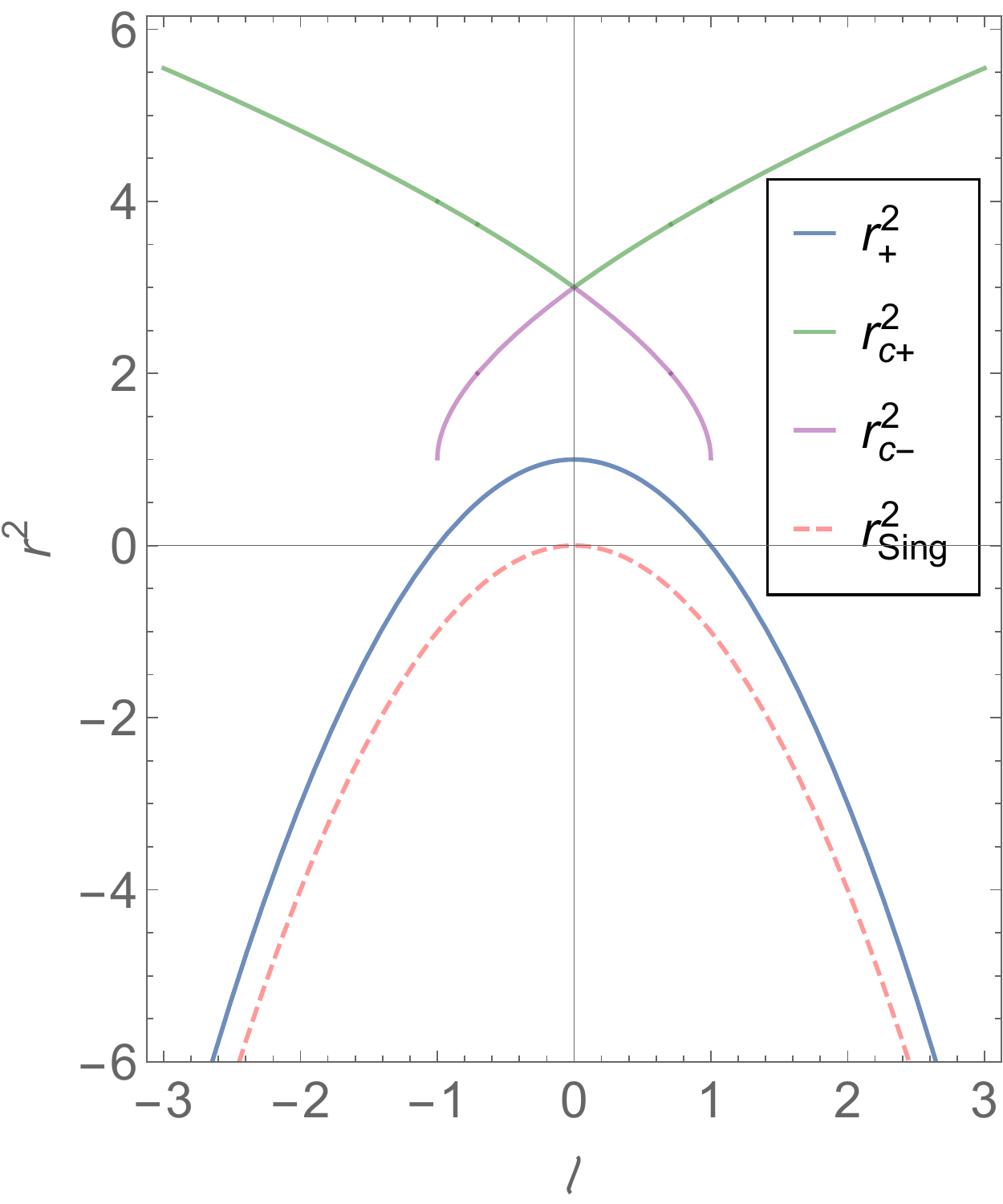}}
\quad
\subfloat[]{\includegraphics[scale=0.37]{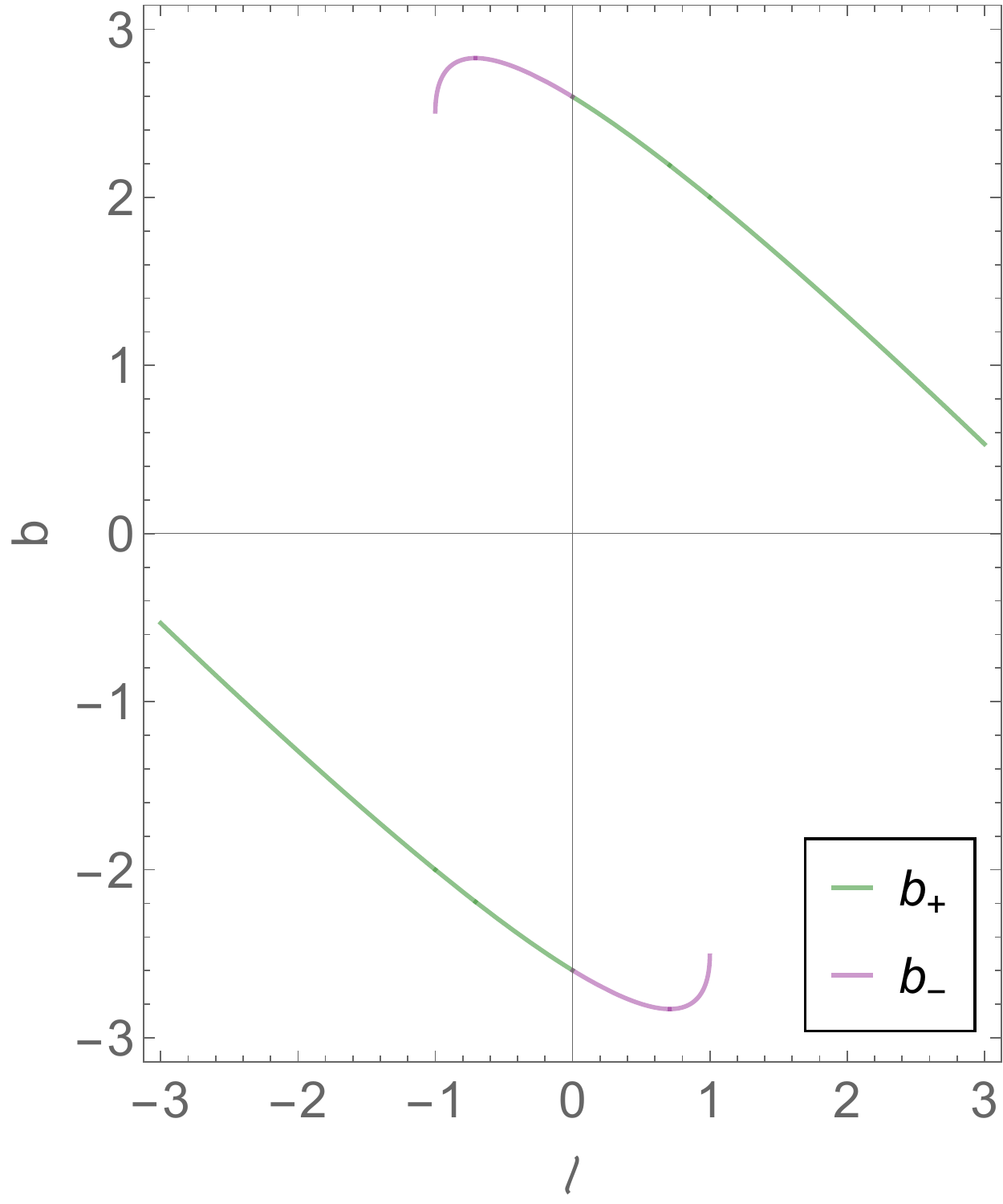}}
\quad
\subfloat[]{\includegraphics[scale=0.37]{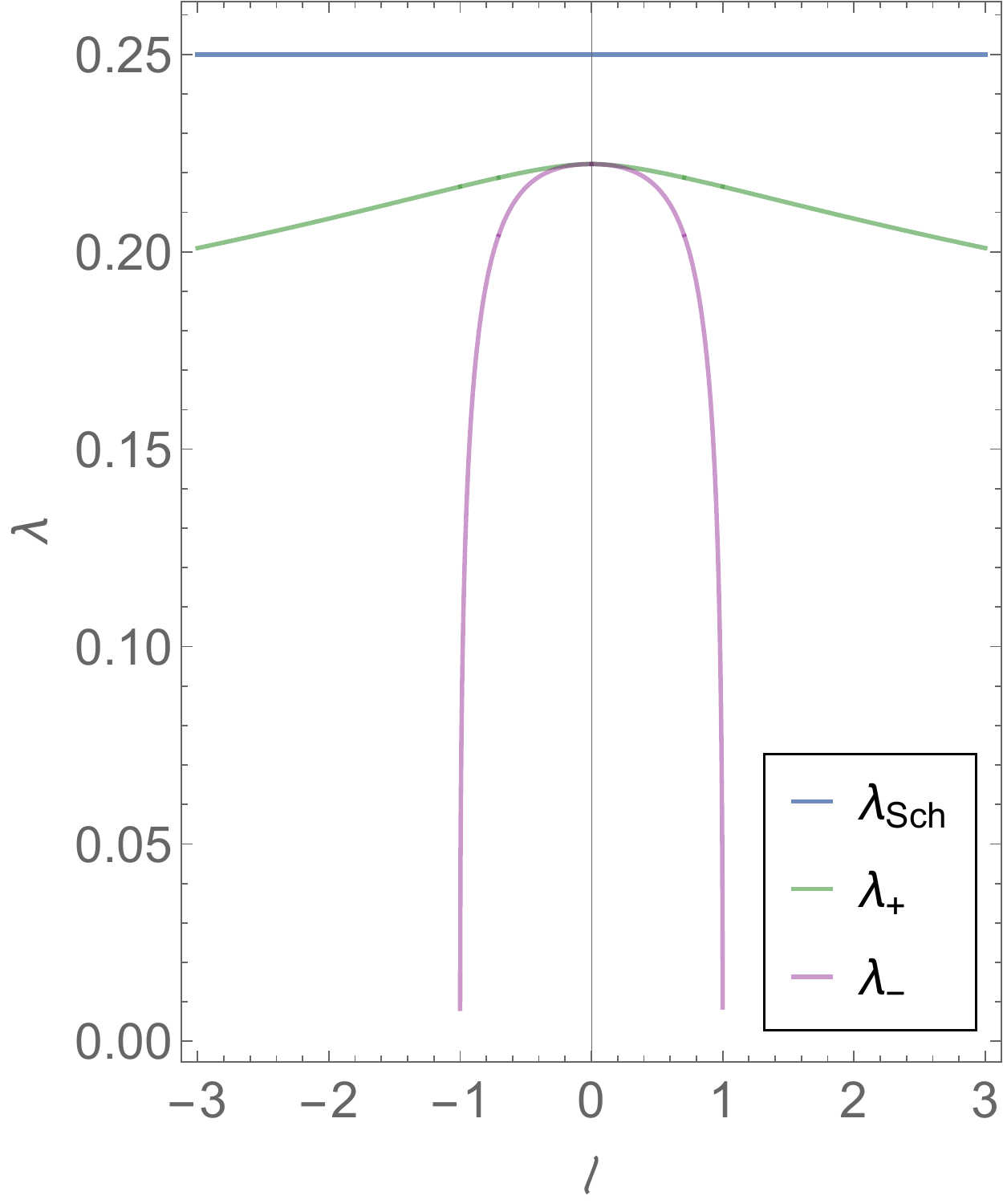}}
\caption{\small{a) In red the maximum values of $r_{\rm sing}^2$, in blue the values of $r_+^2$, In green and purple the values of $r_{c,+}^2$ and $r_{c,-}^2$ respectively. b) In green and purple the values of $b_{c,+}$ and $b_{c,-}$ respectively. c) In green and purple the values of $\lambda_{+}$ and $\lambda_{-}$ respectively. In all plots we set ${\cal M} =1$ and $\ell_1 = 0.8$.}}
\label{plot-BMPV}
\end{figure}
 Extremal supersymmetric  solutions (BMPV geometries) are found for mass, charge and angular momenta \be
{\cal M}={\cal Q} \qquad, \qquad \ell_1=-\ell_2=\ell
\ee 
  For this choice the characteristic polynomials  become
 \begin{equation}
\begin{aligned}
 {\cal R}(x)& =x^3 -(b({\cal M} - x) +  \ell (x-2 {\cal M}) )^2 \\
 {\cal S}_0(x) & =  \frac{x^3+{\cal M} \,x \,\ell\,  (b+\ell) -{\cal M}^2\, \ell \, \left(b+2\ell \right)}{({\cal M}-x)^2}
\end{aligned}
\end{equation}
with
\be
x=r^2+\ell^2=\rho^2
\ee
 The critical equations  ${\cal R}(x_c)= {\cal R}'(x_c)=0$  can be easily solved for $b_{\psi,c}$ and $\ell$ in terms of $x_c$. 
 The two solutions corresponding to the inner and outer radii of the photon sphere will be again denoted by subscripts $+,-$ and can be written as 
 \be
 \begin{aligned}
 b_c &=\pm {\sqrt{x_c}(x_c-6\mathcal{M}) \over 2\mathcal{M}} \qquad ,  \qquad \ell = \mp {\sqrt{x_c}(x_c-3\mathcal{M}) \over 2\mathcal{M}}
  \end{aligned}
 \ee
  We can think of $x_c$ as a parametrization of the angular momentum variable. For the Lyapunov exponent one finds
  \be
  \lambda=  {   \sqrt{ {\cal R}''(x_c)/2 } \over   {\cal S}_0(x_c) }=  {  x_c-{\cal M}\over  \sqrt{3 x_c^3 } }
  \ee
In figure \ref{plot-BMPV}a) we display the locations of $r_{c\pm}$, the outer horizon $r_+$ and the singularity $r_{\rm sing}(0)$. We observe that the outer photon sphere $r_{c+}$ exists always while the inner photon-sphere $r_{c-}$ exists only inside the BH region.  In figure \ref{plot-BMPV}b) and \ref{plot-BMPV}c) we display the critical impact parameters and the Lyapunov exponents. We find that the Lyapunov exponent always decreases monotonically with the angular momenta reaching its maximum for the Schwarzschild solution.  

\subsubsection{General and JMaRT solutions}
\begin{figure}[t]
\centering
\subfloat[]{\includegraphics[scale=0.37]{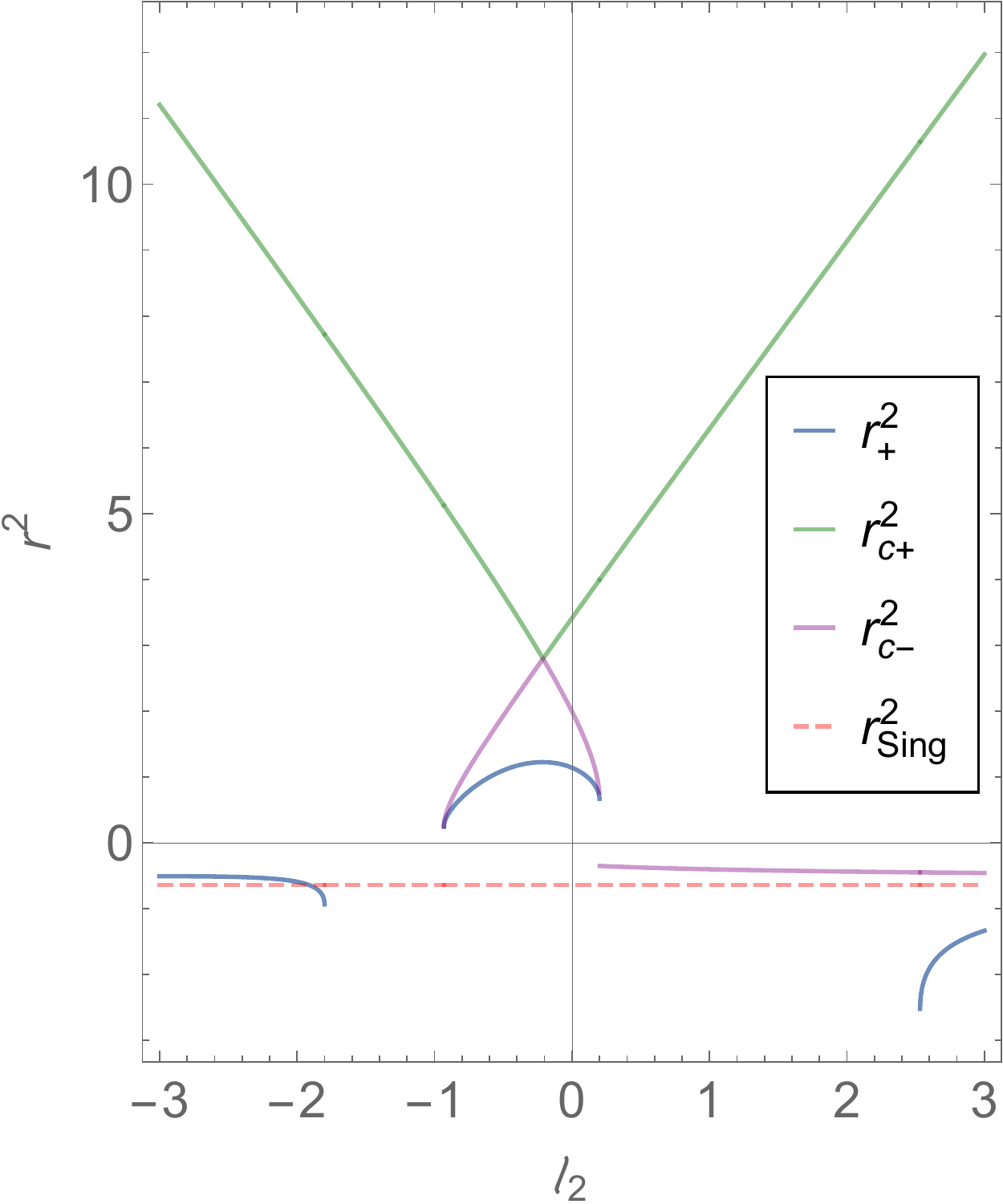}}
\quad
\subfloat[]{\includegraphics[scale=0.37]{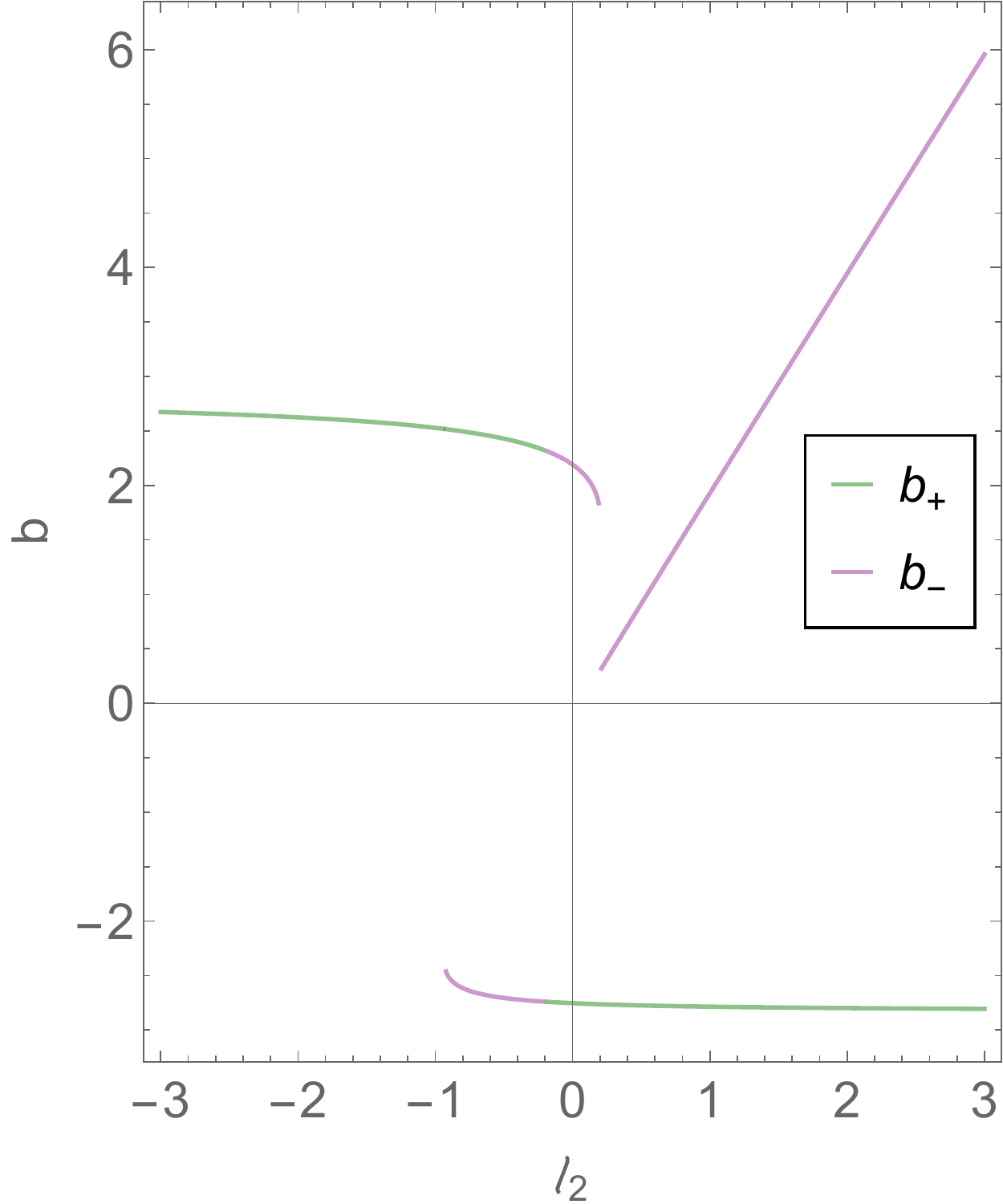}}
\quad
\subfloat[]{\includegraphics[scale=0.37]{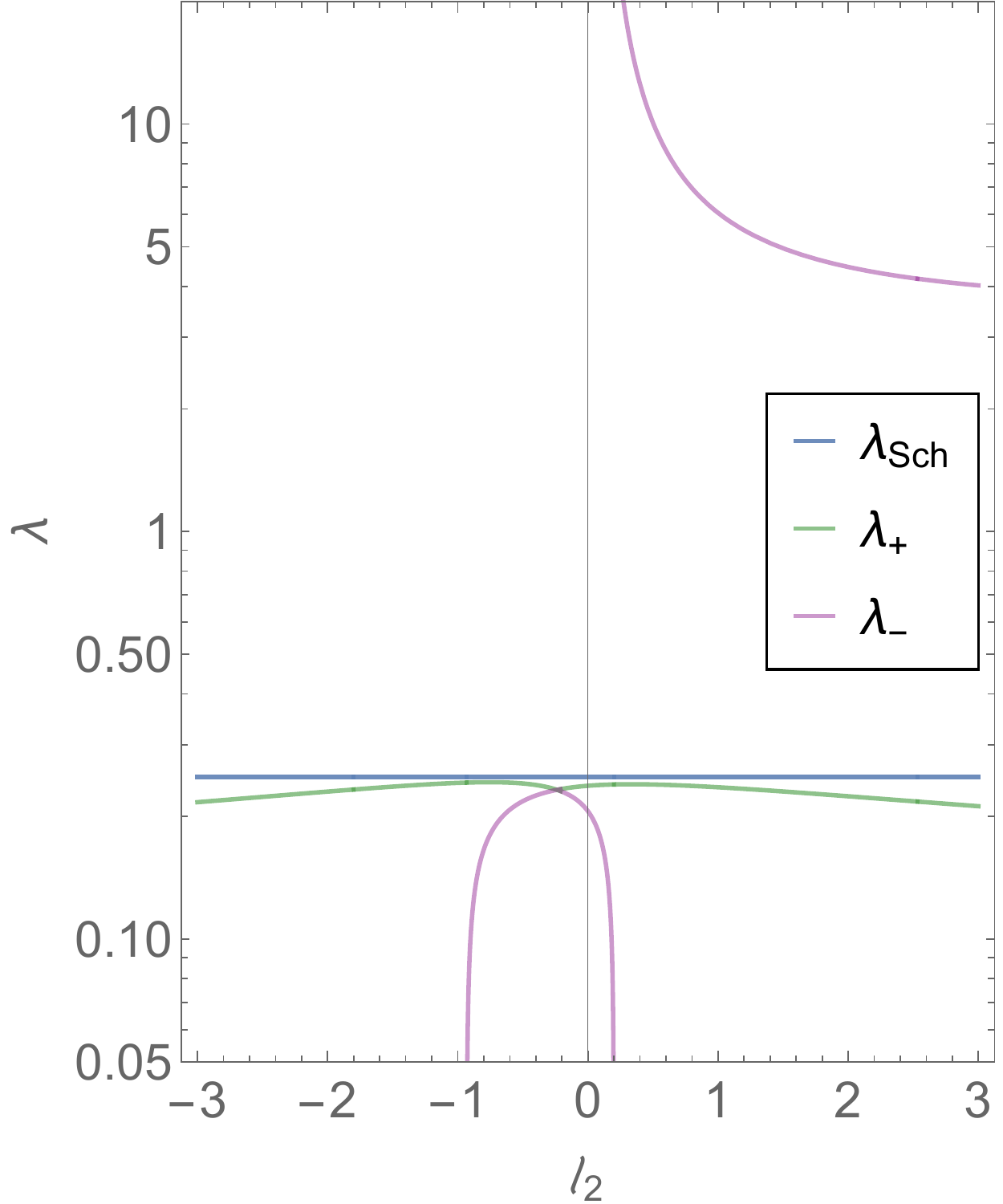}}
\caption{\small{a) In red the maximum values of $r_{\rm sing}^2$, in blue the values of $r_+^2$, In green and purple the values of $r_{c,+}^2$ and $r_{c,-}^2$ respectively. b) In green and purple the values of $b_{c,+}$ and $b_{c,-}$ respectively. c) In green and purple the values of $\lambda_{+}$ and $\lambda_{-}$ respectively. In all plots we set ${\cal M} =1$, ${\cal Q} = 0.5$ and $\ell_1 = 0.8$.}}
\label{plot-CCLP}
\end{figure}

\begin{figure}[t]
\centering
\subfloat[]{\includegraphics[scale=0.7]{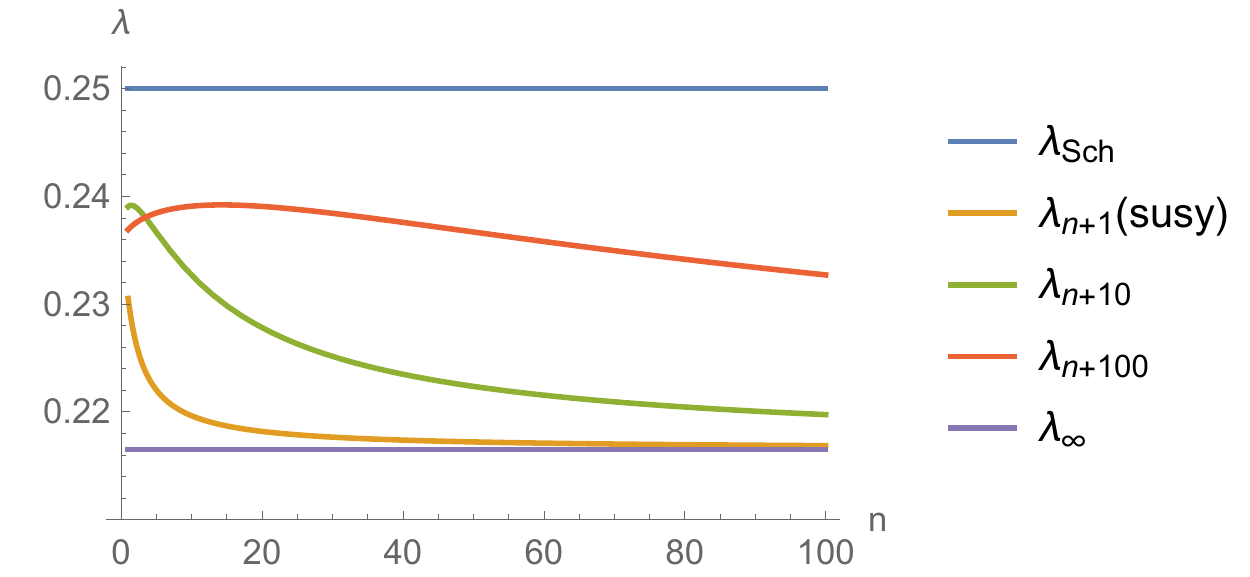}}
\caption{\small{Lyapunov exponents of different JMaRT solutions including the GMS limit $m = n+1$ and its $n\to\infty$ limit. We plot in blue the 5-dimensional Schwarzschild BH Lyapunov exponent.}}
\label{plot_JMaRT_BMPV}
\end{figure}

Finally we consider generic solutions and JMaRT geometries. The photon-sphere in these cases can be found only numerically. In the generic case one finds three different regions see figure \ref{plot_JMaRT_BMPV}a). The BH region is located around the origin (small $\ell_2$) and it is characterized by the existence of an inner and outer photon spheres with Lyapunov exponents always below the bound $\lambda_\pm <\lambda_{\rm Schw}$. Naked singularities (positive and large enough $\ell_2$ ) exhibit an outer and an inner photon sphere, the latter crashing against the singularity and leading to a blow up of the corresponding Lyapunov exponent $\lambda_{+} < \lambda_{\rm Schw} <\lambda_{-}$. Over-rotating geometries  ( negative and large enough $\ell_2$ ) exhibit only an outer photon sphere with Lyapunov exponent always below the bound.

Finally JMaRT geometries are defined by taking ${\cal M}$, ${\cal Q}$, $\ell_1$, $\ell_2$ given by (\ref{mqJMaRT}) and (\ref{ell1_ell2_eqs}). Plugging these formulae into (\ref{poly_coeff}) and (\ref{s0}) and solving the critical equations (\ref{critical}) for $r_c$ and $b_c$ one determines the photon sphere radius and the critical value for the impact parameter. The Lyapunov exponent is then given by eq. (\ref{lambda}). The resulting plots are displayed in figure \ref{plot_JMaRT_BMPV}b) for various families of geometries with ${\cal M}=1$. We observe that the
Lyapunov exponent is always above that of the $m,n\to\infty$ geometry and below the bound set by a Schwarzschild BH of the same mass.

 \section{Conclusions and outlook}
  
We have shown that quite independently of their `intimate' nature, BHs, over-rotating, naked singularities and smooth horizon-less geometries are always surrounded by a photon-sphere (or light-ring), but they can be discriminated from one another by the sizes and shapes of the light rings.  In particular we have found that the Lyapunov exponent $\lambda$ of any smooth or singular geometry with a singularity hidden behind a horizon is always bounded by  $\lambda_{Schw}$, the Lyapunov exponent of the Schwarzschild BH with the same mass ${\cal M}$, while naked singularities violate the bound. The violation of the bound in this pathological case is not surprising since the inner photon sphere 
of these geometries crash against the singularity leading to arbitrary critical impact parameters (small target sizes).  Remarkably, the cosmic censorship \cite{Penrose:1969pc} prevents the violation of the bound.
 
Rotating (non-BPS) BH's and JMaRT solutions are surrounded by an ergo-region where a time-like Killing vector becomes space-like. It would be interesting to study the Penrose process \cite{Penrose:1971uk} in this general context (the JMaRT case has been addressed in \cite{Bianchi:2019lmi}). Other properties that may discriminate between BHs and fuzz-balls are the multi-pole moments (see \cite{Bianchi:2020miz,Bianchi:2020bxa,Bena:2020see,Bena:2020uup} for recent developments in this direction, and \cite{Mayerson:2020tpn} for a nice review on the subject) and the QNM's that we plan to analyse in the near future. in this respect, it is worth recalling that the Lyapunov exponent $\lambda$ sets the scale of the imaginary part of $\omega_{\rm QNM}$, thus motivating our detailed analysis of near critical null geodesics in the vicinity of the photon-sphere.

We would like to conclude with a word or two of caution. As already mentioned, when we talk about naked singularities, that are known to be forbidden by cosmic censorship \cite{Penrose:1969pc}, we don't want to propose them as a possibility but rather suggest that 
some general properties of gravity like the existence of a photon-sphere extend also to these pathological cases. We want to stress that in a gravity theory with matter the existence of a naked singularity by itself cannot be excluded if this is accompanied with a blow up of matter fields signaling a breaking of the effective picture and the emerging of higher dimensions.  This is the case of JMaRT solutions which cannot be thought as a {\it bona fide} micro-states since they do not match the asymptotics for a BH solution to exist.  Both JMaRT solutions and their BPS limit, represented by the GMS solutions, may be thought of as `fake' fuzz-balls in that one cannot tune any parameter in the solution so that it matches a BH. Yet, the latter achieve the goal for very large angular momenta. Finding {\it bona fide} `generic' micro-state geometries of rotating (BPS) BH's turns out to be challenging both in four and five-dimensions \cite{Giusto:2004ip,Bena:2005va, Berglund:2005vb, Saxena:2005uk,Bena:2006kb,Bena:2007kg,Bena:2007qc,Giusto:2009qq,Giusto:2011fy,Lunin:2012gp,Giusto:2013bda,Gibbons:2013tqa,Bena:2015bea,Lunin:2015hma,Bena:2016agb, Bena:2016ypk,Pieri:2016cqz, Bianchi:2016bgx, Pieri:2016pdt, Bianchi:2017bxl}. Exploring their properties and identifying their characteristic features remain an interesting issue worth pursuing in the future.

\appendix

\section{JMaRT/CCLP dictionary}
\label{sectionJMaRT}

In this section we present the dictionary between CCLP and JMaRT variables.

\subsection{The general six-dimensional D1-D5-p solution }

 The general ten-dimensional JMaRT geometries  can be obtained starting from a rotating BH in five dimensions (with mass $M$ and rotation parameters $a_1$, $a_2$) lifted to ten dimensions by adding a five-torus $T^4\times S^1$. The D1, D5, and p charges are then generated by a sequence of three boots  $\Lambda_1$,  $\Lambda_5$,  $\Lambda_p$ with boost parameters  $\delta_1$,  $\delta_5$,  $\delta_p$ along the $S^1$-circle intercalated by S and T-dualities \cite{Giusto:2004id,Giusto:2004kj}.
 
Setting all charges to be equal $Q_1=Q_5=Q_p={\cal Q}$, i.e. $\delta_1=\delta_5=\delta_p$, the dilaton field and the metric of $T^4$ become constant, so the solution can be viewed as a solution of a six-dimensional gravity coupled to an anti-self dual 2-form (\textit{i.e.} with $H_3=-*H_3$).
 
 The six-dimensional metric can be written as
\begin{equation}
\begin{aligned}\label{gms}
ds_2^6 &=  ds_5^2 +\left(dy-{A\over \sqrt{3}} \right)^2\\
&{\rm where}\\
&ds_5^2 = -\left(1-\frac{M}{f}\right)Z^{-2}\left(dt + \alpha\right)^2 + Z ds_4^2
\\
&A = {\sqrt{3}\mathcal{Q}\over Z \, f} 
(dt -\sinh\delta\, \alpha_1+  \cosh\delta \, \alpha_2 )
\end{aligned}
\end{equation}
with  
\be
\begin{aligned}
 Z &= 1 + \frac{ M\sinh^2\delta}{f}
\quad
,
\quad
f = \hat r^2 + a_1^2\sin^2\theta + a_2^2\cos^2\theta 
\quad
,
\quad
Z f = \rho^2\\
\alpha  &= \frac{M \sinh^3\delta  }{f}\alpha_1-\frac{M \cosh^3\delta }{f-M}\alpha_2\\
\alpha_1 &=  a_1 \, \sin^2\theta \,d\phi + a_2 \cos^2\theta \,d\psi \qquad , \qquad 
\alpha_2 =  a_2 \, \sin^2\theta \,d\phi + a_1 \cos^2\theta \,d\psi \\
\end{aligned}
\ee
and 
\be
\begin{aligned}
ds_4^2 =& f\left[\frac{ \hat r^2 d \hat r^2}{\left( \hat r^2 + a_1^2\right) \left(\hat r^2 + a_2^2\right)-M \hat r^2}+d\theta^2\right] + \frac{M  }{f-M}\alpha_2^2\\
&+( \hat r^2 +a_1^2)\,\cos^2\theta\,d\psi^2 + ( \hat r^2 +a_2^2)\,\sin^2\theta \, d\phi^2
\end{aligned}
\ee
The metric  (\ref{gms}) matches  (\ref{metric6d})    after the identifications
\be
\begin{aligned}\label{dict}
\ell_1 & = a_1 \sinh\delta -a_2 \cosh\delta \quad, \quad \ell_2=-a_1 \cosh\delta +a_2 \sinh\delta\,,\\
{\cal M} &= {M\over 2} \cosh2\delta \,,\qquad {\cal Q}={M\over 2} \sinh 2\delta\,,\\
r^2 &= \hat r^2+ a_1 a_2 \sinh2\delta +(M-a_1^2-a_2^2)\sinh^2\delta 
\end{aligned}
\ee
in this notation the angular momenta of the 5d-solution read
\begin{equation}
\begin{aligned}
   J_\psi 
={M \over 3} \left(a_2 \sinh^3\delta- a_1 \cosh ^3\delta \right)  \quad, \quad 
  J_\phi= {M \over 3} \left(a_1 \sinh^3\delta-a_2 \cosh ^3\delta\right) \end{aligned}
\end{equation}

\subsection{JMaRT geometries}
 
 Smooth horizon-less geometries obtain by choosing the mass, boost and angular parameters $M$, $\delta$, $a_1$, $a_2$, and the
 radius $R$ of the $y$-circle such that  
 \begin{equation}
\begin{aligned}\label{JMaRTeq}
M &= a_1^2+a_2^2-a_1 a_2 (\coth[\delta]^3+\tanh[\delta]^3)  \\
 R &=  {M \sqrt{\cosh\delta \sinh\delta} \over \sqrt{a_1a_2} (\coth\delta^3-\tanh\delta^3)}    \\
 \alpha R &={\sinh(2\delta) R \over 2(a_2 \sinh^3\delta-a_1 \cosh^3\delta)} =-n  \\
     \beta R & ={\sinh(2\delta) R \over 2(a_1 \sinh^3\delta-a_2 \cosh^3\delta)} =m
\end{aligned}
\end{equation} 
 Using the dictionary (\ref{dict}) one can check equivalence with the JMaRT defining equations (\ref{mqJMaRT}) , (\ref{y-circle_radius}), (\ref{quantization_rules}) and (\ref{JMaRTeq}).

\subsection{Extremal limit }

The extremal limit of the JMaRT solution obtains by sending $M \to 0$, $\delta \to \infty$ with $M e^\delta$ finite. In this limit the mass ${\cal M}$, charge ${\cal Q}$ and angular momenta $J_\psi$, $J_\phi$ (as well as $\ell_1$, $\ell_2$) stay finite.
 More precisely, one writes
\begin{equation}
\begin{aligned}
a_1 &\simeq -\frac{\left(J_\psi+J_\phi\right)}{3 \sqrt{   M {\cal Q}  }}
 +\frac{\left(J_\phi- J_\psi\right) \sqrt{ M}}{4 {\cal Q}^{3\over 2} }+\ldots \\
a_2 & \simeq -\frac{\left(J_\psi+J _\phi \right)}{3 \sqrt{ M   {\cal Q}}}
 +\frac{\left(J_\psi-J_\phi\right) \sqrt{ M}}{4 {\cal Q}^{3\over 2} }+\ldots \\
 \hat r^2 &= \check{r}^2-\frac{(J_\psi + J_\phi)^2}{9 M {\cal Q} } + a_\phi^2 - a_\psi^2
\end{aligned}
\end{equation}
with
\begin{equation}
\begin{aligned}
{\cal Q}^2 \, a_{\phi }^2&=\frac{1}{6} \left(J _\psi+J_\phi \right) \left(J_\phi-J_\psi+\sqrt{\left(J_\psi-J_\phi\right)^2-4 {\cal Q}^3}\right)
\\
{\cal Q}^2 \, a_{\psi }^2&=\frac{1}{6} \left(J _\psi+J_\phi \right) \left(J_\phi-J_\psi-\sqrt{\left(J_\psi-J_\phi\right)^2-4 {\cal Q}^3}\right)\end{aligned}
\end{equation}
In this limit, one finds that $ds_5^2$ and $A$ in (\ref{metric6d}) reduce to the five-dimensional extremal metric and gauge fields in (\ref{cclpext}).

In terms of these variables the five-dimensional metric and  the gauge field $A$ can be written as
\begin{equation}
\begin{aligned} \label{cclpext}
ds_5^2 &= -Z^{-2}\left(dt + \alpha\right)^2 + Z ds_{4, {\rm ext} }^2
\\
A &= \frac{{\cal Q}}{Z \, f}\left[ - dt +\frac{(a_{\psi }^2+a_{\phi }^2)}{2 a_{\psi } a_{\phi }}
 \, \alpha_- +{a_{\phi } \,a_{\psi } \over 2 {\cal Q}}\, 
\alpha_+ \right]    
\end{aligned}
\end{equation}
where
\bea
\label{effeZeta}
f &=& \check{r}^2 + a_\phi^2\cos^2\theta - a_\psi^2\sin^2\theta
\quad , \quad
Z = 1 + \frac{{\cal Q}}{f}. \\
\alpha &=& \frac{1}{2  f} \left[ { {\cal Q} (a_{\psi }^2+a_{\phi }^2) \over a_{\psi } a_{\phi } }
\alpha_- +(1 + 2 Z) a_{\psi } a_{\phi }   \alpha_+ \right] 
\quad , \quad \alpha_\pm  = \sqrt{\cal Q} (d\phi \sin^2\theta \pm d\psi \cos^2\theta)  \nn
\eea
and  
\begin{equation}
\label{ds4extr}
\begin{aligned}
ds_{4, {\rm ext} }^2  =& f\left(\frac{d \check{r}^2}{ \check{r}^2 + a_\phi^2 - a_\psi^2}+d\theta^2\right)
+
\frac{a_\phi^2 a_\psi^2}{f}\left(d\psi\cos^2\theta + d\phi\sin^2\theta\right)^2
\\
&+
d\psi^2 \cos^2\theta\left(\check{r}^2-a_\psi^2\right)
+
d\phi^2 \sin^2\theta\left( \check{r}^2 + a_\phi^2\right)
\end{aligned}
\end{equation}
Note that the 5-d geometry is a (pseudo)riemannian fibration over a 4-d base whose metric $ds_{4, {\rm ext} }^2$ is hyper-kahlerian (Ricci flat).

The extremal (in general non supersymmetric) solution can be obtained by setting ${\cal M}= {\cal Q}$ in  (\ref{CCLPmetric}) and keeping arbitrary $\ell_1$, $\ell_2$.   In this case, it is convenient to introduce the radial coordinate $\check{r}$ and the  angular parameters $a_\phi$ and $a_\psi$  defined in terms of $r$, $\ell_1$ and $\ell_2$ via
\begin{equation}
\begin{aligned}
\ell_1 &= \frac{{\cal Q} \left(a_{\psi }^2+a_{\phi }^2\right)+a_{\psi }^2 a_{\phi }^2}{2 \sqrt{{\cal Q}} a_{\psi } a_{\phi }}
\quad
,
\quad
\ell_2 = -\frac{{\cal Q} \left(a_{\psi }^2+a_{\phi }^2\right)-a_{\psi }^2 a_{\phi }^2}{2 \sqrt{{\cal Q}} a_{\psi } a_{\phi }} \\
r^2 &= \check{r}^2 - \frac{\left[{\cal Q} (a_{\psi }^2-a_{\phi }^2)+a_{\psi }^2 a_{\phi }^2\right]^2}{4 {\cal Q} a_{\psi }^2 a_{\phi }^2}
\end{aligned}
\end{equation}

\bibliographystyle{JHEP}
\bibliography{lightring5d}
\end{document}